\documentclass[11pt]{article}
\usepackage[margin=1in]{geometry}
\usepackage{amsmath,amssymb,amsthm,mathtools,bm,comment,tikz}
\usepackage[round,authoryear]{natbib}
\usepackage[hypertexnames=false]{hyperref}
\usepackage{enumitem}
\usepackage{float}
\usepackage{graphicx}
\usepackage{xcolor}
\usepackage{booktabs,multirow}

\hypersetup{colorlinks=true,linkcolor=blue,citecolor=blue,urlcolor=blue}

\newtheorem{assumption}{Assumption}

\newtheorem{theorem}{Theorem}
\newtheorem{proposition}{Proposition}
\newtheorem{lemma}{Lemma}
\newtheorem{remark}{Remark}

\newcommand{\R}{\mathbb{R}}
\newcommand{\E}{\mathbb{E}}
\newcommand{\PP}{\mathbb{P}}
\newcommand{\Var}{\operatorname{Var}}

\newcommand{\FE}{\mathrm{FE}}
\newcommand{\RE}{\mathrm{RE}}

\newcommand{\cD}{\mathcal{D}}

\title{Bias-aware meta-analysis: beyond fixed-effect and random-effects meta-analyses}
\author{Mingxuan Zou\thanks{Department of Statistics, University of Chicago, Chicago, IL 60637, USA; e-mail: \texttt{mingxuanzou@uchicago.edu}}
 \and Peng Ding\thanks{Department of Statistics, University of California, Berkeley, CA 94720, USA; e-mail: \texttt{pengdingpku@berkeley.edu}}}
\date{\today}

\begin{document}
\maketitle

\begin{abstract}
    
Meta-analysis aims to synthesize multiple studies to obtain a summary estimate of a population parameter of interest. However, existing meta-analysis methods require strong assumptions that render them not robust to bias and heterogeneity of the studies. In particular, fixed-effect meta-analysis assumes a common parameter across the studies, whereas random-effects meta-analysis models heterogeneity as independent and identically distributed random effects, which still implies a common parameter marginally over the random effects. Due to their limitations, we propose a bias-aware meta-analysis framework that allows for heterogeneous studies with fixed, unknown biases. With no bias, our bias-aware meta-analysis reduces to fixed-effect meta-analysis; with a given bound on the  bias, we can still use the fixed-effect meta-analysis point estimator and standard error, but we must inflate the critical value for confidence intervals from the standard normal quantile to a folded normal quantile. We recommend reporting the {\it maximum bias for significance}, the maximum value of the bias such that the confidence intervals exclude the null value. We propose a formal model checking procedure and recommend reporting the {\it minimum bias for compatibility}, the minimum value of the bias such that the bias-aware meta-analysis model is compatible with the observed data. We  discuss the practical implementation of our method and, in particular, the approaches to calibrating the maximum bias for significance.  We also compare our method with  traditional fixed-effect and random-effects meta-analyses. 

\end{abstract}


\section{Introduction to meta-analysis}

\subsection{Literature review and critiques}

Meta-analysis, coined by \citet{glass1976}, aims to combine statistical evidence from multiple studies. It has a long history in statistics.
\citet{fisher1932} and \citet{pearson1934} proposed methods to combine independent $p$-values (see also \citet{owen2009}). \citet{fisher1932} proposed a method to combine independent estimates by weighting them based on the inverse of their variances, which is essentially the fixed-effect meta-analysis (FEMA) in the modern literature. Motivated by \citet{fisher1935design}'s analysis of multiple agricultural experiments, \citet{yatescochran1938} proposed a framework for meta-analysis that allows for between-study heterogeneity, and introduced many key concepts (e.g., Cochran's $Q$-statistic) that appeared later in \citet{cochran1954}'s development for combining estimates from multiple experiments. Building upon \citet{cochran1954}, \citet{dersimonian1986} provided a canonical presentation of the random-effects meta-analysis (REMA), which contained a parameter to summarize the treatment effects across studies as well as a parameter to quantify the between-study heterogeneity. REMA is widely used in modern medical research, and \citet{dersimonian1986} is one of the most cited papers in statistics \citep{laird2014,dersimonian2015}.

The key issue of meta-analysis is the potential bias and heterogeneity of the studies, as clearly articulated in \citet{yatescochran1938}, \citet{cochran1954}, and \citet{dersimonian1986}. 
The traditional FEMA and REMA impose strong assumptions on the bias and heterogeneity. 
Under the FEMA model, all studies provide an unbiased estimate for a common effect, which is the ``fixed effect.'' Therefore, the FEMA estimator equals the weighted estimator based on the inverse of the variances of the study-specific estimates. It is the maximum likelihood estimate and enjoys many optimality properties. However, the FEMA model imposes the strong assumption that all studies have zero bias and are homogeneous. In fact, it can be rejected by the observed data based on Cochran's $Q$-test \citep{cochran1954}. 
The REMA model can address the limitation of the FEMA model in some sense. 
Under the REMA model, each study can have its own parameter, and the study-specific parameters are drawn from a random variable. Therefore, the study-specific parameters are the ``random effects.'' The inferential targets under REMA are the mean and variance of the random effects, with the mean providing a summary of the treatment effect and the variance quantifying the heterogeneity across studies. However, the REMA model still imposes the strong assumption that each study estimate is unbiased for the target effect, marginally over the study-specific random effects. That is, while the REMA model allows for certain heterogeneity, it imposes a parametric structure on it. 
The REMA model cannot capture the intrinsic heterogeneity when the study-specific estimates are not centered at the target effect.

\subsection{Our contributions and paper organization}

We depart from the existing literature of FEMA and REMA by providing an alternative formulation of the bias and heterogeneity. 
We propose a bias-aware meta-analysis (BAMA) framework that allows for fixed, study-specific bias across studies. Importantly, we do not impose any distributional assumptions on the biases and assume they are unknown nuisance parameters.

Without any assumptions on the study-specific bias, we cannot make nontrivial inference. To make meaningful progress, we impose an upper bound on the maximum standardized bias, that is, the maximum of the study-specific bias divided by the square root of the harmonic mean of the study-specific variances. Given this bound, we show that we can retain the point estimate and standard error under FEMA but replace the standard-normal critical
value with a folded-normal quantile to construct confidence
intervals under BAMA. The folded-normal distribution has a non-zero center, which depends on the maximum standardized bias. Practitioners can treat the maximum standardized bias as a sensitivity parameter and report a sequence of confidence intervals. This approach has been standard in causal inference to assess the impact of unmeasured confounding \citep[e.g.,][]{rosenbaum1983assessing}.

Based on the above confidence intervals, we propose reporting the {\it maximum bias for significance}, the threshold of the bias beyond which the confidence intervals under BAMA include the null value of the target effect. 
It quantifies how large the bias must be to overturn the significance of the result. 
This notion is related to the b-value in \citet{lin2026}, which was motivated by the E-value in sensitivity analysis in causal inference \citep{vanderweele2017evalue}.  
Practitioners can read off this value from the plot of the confidence sequences against the bias. We further derive an approximate formula for it based on asymptotic analysis, which equals the absolute value of the $t$-statistic under FEMA divided by the square root of the number of studies.

Our BAMA model reduces to the traditional FEMA model when the bias equals zero. Since the FEMA model can be rejected by the observed data based on Cochran's $Q$-test, it is intuitive that our BAMA model can also be rejected if the specified bias parameter is too small to be compatible with the observed data. Under BAMA, we use Cochran's $Q$-statistic for model checking, and show that we must use quantiles of a noncentral chi-squared distribution to compute the $p$-value. We derive the explicit formula of the noncentrality parameter under BAMA. Practitioners can report a sequence of $p$-values for model checking of BAMA at different levels of bias.

Based on the above $p$-values, we propose reporting the {\it minimum bias for compatibility}, the threshold of the bias below which the BAMA model is rejected. It quantifies how large the bias must be to ensure the BAMA model is compatible with the observed data. 
Practitioners can read off this value from the plot of the $p$-values against the bias. We further derive an approximate formula for it based on asymptotic analysis, which equals the square root of the ratio of the between-study variance estimate under REMA to the harmonic mean of the within-study variances.

The minimum bias for compatibility and the maximum bias for significance form an interval. 
When this interval is empty, we declare a null result because there are no bias values to ensure simultaneously that the BAMA model is compatible with the observed data and the confidence intervals under BAMA exclude the null value. We provide characterizations of this regime of null results. 
When this interval is nonempty, there are bias values under which the BAMA model is not rejected and the confidence intervals under BAMA exclude the null value. While this is often the scientifically more interesting regime, we must calibrate the maximum bias for significance to ensure that it is not beyond the plausible range of the bias. We also provide guidance for this calibration.

Overall, our BAMA framework goes beyond the traditional FEMA and REMA. We recommend that practitioners report not only the results under 
FEMA and REMA but also the two key quantities of the minimum bias for compatibility and the maximum bias for significance under BAMA. These together can provide a more complete picture of the problem of meta-analysis.

The rest of the paper proceeds as follows. Section~\ref{sec:background} reviews and
criticizes FEMA and REMA models and also introduces the basic notation for the whole paper.
Section~\ref{sec:BAMA} proposes the BAMA model, discusses the associated inference problem, and introduces the key notion of maximum bias for significance.
Section~\ref{sec:model-checking} discusses model checking of BAMA and introduces the key notion of minimum bias for compatibility.
Section~\ref{sec:recommendations} provides practical recommendations for implementing and interpreting our method. 
Section~\ref{sec:examples} uses numerical examples to illustrate our method. 
Section~\ref{sec:discussion} concludes with additional discussion. The online supplementary material contains additional results and the proofs.

\section{FEMA and REMA: review and critiques}
\label{sec:background}
This section reviews the FEMA and REMA models. The review helps to introduce the basic notation and the motivation for our BAMA model in Section~\ref{sec:BAMA}.

\subsection{FEMA}
\label{sec:FEMA}
\begin{assumption}[FEMA model]
\label{ass:FEMA-model}
Consider $J$ independent studies. In study $j$ $(j=1, \ldots, J)$, let
$\hat\tau_j$ be an effect estimate with known variance $\sigma_j^2$, and assume
\[
\hat\tau_j \sim N(\tau,\sigma_j^2),
\]
for a common fixed effect $\tau$.
\end{assumption}
The FEMA model assumes that all studies estimate the same underlying
effect \(\tau\). The large-sample approximation for each study can justify the normality assumption in Assumption~\ref{ass:FEMA-model}. In many applications, the study-specific estimate
$\hat\tau_j$ satisfies a central limit theorem, and we can consistently estimate its variance.  Therefore, we follow the convention to let $\sigma_j^2$ denote the
corresponding asymptotic variance estimate and write the sampling distribution as normal.  

Under Assumption~\ref{ass:FEMA-model}, the maximum likelihood estimator for $\tau$ is the precision weighting estimator
\begin{equation}
\label{eq:fixed-effect-estimator}
    \hat\tau_{\FE}
    =\sum_{j=1}^J \lambda_{j,\FE}\hat\tau_j,
\end{equation}
with
\[
    \lambda_{j,\FE}=\frac{\sigma_j^{-2}}{\sum_{j=1}^J\sigma_j^{-2}}.
\]
It follows the normal distribution $\hat\tau_{\FE}\sim N(\tau,\sigma_{\FE}^2)$,
with unknown mean $\tau$ and known variance
\[
\sigma_{\FE}^2=\frac{1}{\sum_{j=1}^J\sigma_j^{-2}}.
\]
Hence, under the FEMA model, we can use the Wald-type confidence interval based on $\hat\tau_{\FE}$ and $\sigma_{\FE}^2$ for the common effect $\tau$.
\subsection{REMA}
\label{sec:REMA}
\begin{assumption}[REMA model]
\label{ass:REMA-model}
Consider \(J\) independent studies. In study \(j\) $(j=1,\ldots,J)$, let \(\hat\tau_j\) be the effect estimate with known variance \(\sigma_j^2\), and assume
\[
    \hat\tau_j\mid\tau_j\sim N(\tau_j,\sigma_j^2)
\]
with study-specific effects \(\tau_j\). Moreover, assume $\tau_j$ has the following first two moments: 
\[
    \E(\tau_j)=\tau,
    \qquad
    \Var(\tau_j)=\sigma_{\textup{B}}^2,
\]
with unknown mean $\tau$ and between-study variance \(\sigma_{\textup{B}}^2\).
\end{assumption}

The REMA model assumes that each study has its own random effect \(\tau_j\). Under Assumption~\ref{ass:REMA-model}, these study-specific effects are independent with a common mean and variance, but need not be identically distributed or normal. Thus, REMA changes the target of inference from a common fixed effect that all studies share to the population mean \(\tau\).

Under Assumption~\ref{ass:REMA-model}, marginally, $\widehat\tau_j$ is unbiased for $\tau$ with variance $\sigma_j^2+\sigma_{\textup{B}}^2$.
If \(\sigma_{\textup{B}}^2\) were known, we could use the precision-weighted average with weight \(1/(\sigma_j^2+\sigma_{\textup{B}}^2)\). Building upon \citet{cochran1954}, \citet{dersimonian1986} proposed first estimating \(\sigma_{\textup{B}}^2\) and
then plugging the estimate into these precision weights. We review their estimator below.

Under Assumption~\ref{ass:REMA-model}, Cochran's \(Q\) statistic
\citep{cochran1954} 
\[
    Q
    =
    \sum_{j=1}^J \sigma_j^{-2}
    (\hat\tau_j-\hat\tau_{\FE})^2
\]
has mean
\begin{equation}
\label{eq:expectation-of-Q_FE}
    \E(Q)
    =
    J-1
    +
    \left(
        \sum_{j=1}^J \sigma_j^{-2}
        -
        \frac{\sum_{j=1}^J \sigma_j^{-4}}
        {\sum_{j=1}^J \sigma_j^{-2}}
    \right)\sigma_{\textup{B}}^2,
\end{equation}
which motivates the estimator
\begin{equation}
\label{eq:hatsigmaB}
    \hat\sigma_{\textup{B}}^2
    =
    \left(
            \sum_{j=1}^J \sigma_j^{-2}
            -
            \frac{\sum_{j=1}^J \sigma_j^{-4}}
            {\sum_{j=1}^J \sigma_j^{-2}}\right)^{-1}
            \left(Q-J+1\right)_+,
\end{equation}
where \((a)_+=\max\{0,a\}\). The corresponding estimator for $\tau$ is
\[
    \hat\tau_{\RE}
    =
    \sum_{j=1}^J \hat\lambda_{j,\RE}\hat\tau_j,
\]
with
\[
    \hat\lambda_{j,\RE}
    =
    \frac{(\sigma_j^2+\hat\sigma_{\textup{B}}^2)^{-1}}
    {\sum_{k=1}^J(\sigma_k^2+\hat\sigma_{\textup{B}}^2)^{-1}}.
\]
It has the following variance estimator
\[
    \hat\sigma_{\RE}^2
    =
    \left\{
        \sum_{j=1}^J
        \left(\sigma_j^2+\hat\sigma_{\textup{B}}^2\right)^{-1}
    \right\}^{-1}.
\]

Hence, under the REMA model, we can use the Wald-type confidence interval based on
$\hat\tau_{\RE}$ and $\hat\sigma_{\RE}^2$ for the mean $\tau$. Although this is a standard
procedure, we do not find its rigorous asymptotic theory in the literature and thus provide a
proof in Appendix~\ref{app:standard-relaxations}.

\begin{remark}[FEMA estimator under the REMA model]
\label{rem:FEMA-under-REMA}
We can also use the FEMA estimator \(\hat\tau_{\FE}\) under the
REMA model, but we must inflate its variance estimator.  
Although this procedure is not standard and not widely used, we view it as a competitor to the preceding inference procedure because of its simplicity and its connection to our BAMA model below. The point estimator $\hat\tau_{\FE}$ remains unchanged, and we must inflate its variance to account for heterogeneity.
Under Assumption~\ref{ass:REMA-model}, $    \hat\tau_{\FE}$ has mean $\tau$ and variance: 
\begin{equation}
\label{eq:FEMA-variance-under-REMA}
 \sigma_{\FE\dagger}^2 =    \sigma_{\FE}^2
        +\sigma_{\textup{B}}^2
        \frac{\sum_{j=1}^J\sigma_j^{-4}}
        {\left(\sum_{j=1}^J\sigma_j^{-2}\right)^2}.
\end{equation}
The second term in \eqref{eq:FEMA-variance-under-REMA} gives the inflation due to the between-study heterogeneity. We can estimate $\sigma_{\textup{B}}^2$ by $\hat\sigma_{\textup{B}}^2$ and estimate the inflated variance by replacing $\sigma_{\textup{B}}^2$ in \eqref{eq:FEMA-variance-under-REMA} with $\hat\sigma_{\textup{B}}^2$: 
\begin{equation}
\label{eq::FEMA-varianceestimate-under-REMA}
\hat{\sigma}_{\FE\dagger}^2 = \sigma_{\FE}^2
        + \hat\sigma_{\textup{B}}^2
        \frac{\sum_{j=1}^J\sigma_j^{-4}}
        {\left(\sum_{j=1}^J\sigma_j^{-2}\right)^2}.
\end{equation}

Hence, under the REMA model, we can use the Wald-type confidence interval based on $\hat\tau_{\FE}$ and $\hat{\sigma}_{\FE\dagger}^2$. We provide a rigorous proof in Appendix~\ref{app:standard-relaxations}.
\citet{higgins2009} mentioned this method. 
\citet{henmicopas2010} discussed inference based on $\hat\tau_{\FE}$ and $\hat{\sigma}_{\FE\dagger}^2$ with a normality assumption on study-specific effects $\tau_j$'s, but proposed a finer critical value that accounts for uncertainty in estimating $\sigma_{\textup{B}}^2$. 
\end{remark}

\begin{remark}[The REMA model with normal random effects]
It is also standard to see REMA with a stronger normality assumption on the $\tau_j$'s. The normality assumption on the \(\hat\tau_j\mid\tau_j\)'s is justified by central limit theorems, whereas the normality assumption on the $\tau_j$'s is not. See \citet{normand1999meta} and \citet{henmicopas2010} for reviews of methods under the stronger REMA model. We focus on the REMA model in Assumption \ref{ass:REMA-model}. 
\end{remark}

\section{BAMA}
\label{sec:BAMA}
The FEMA model in Section~\ref{sec:FEMA} assumes away heterogeneity across studies.
The REMA model in Section~\ref{sec:REMA} allows for heterogeneity but treats the study-specific
effects $\tau_j$ as independent random variables with a common mean and variance. Both the FEMA and REMA models assume a common parameter $\tau$ across studies, marginally over the random effects. 
However, studies may differ systematically in their populations, designs, or
treatment implementations, so both FEMA and REMA models are too strong.
In this section, we propose a different meta-analysis formulation that makes no distributional assumptions on the study-specific effects and discuss the inference issues from a sensitivity analysis perspective. We call it BAMA for ``bias-aware meta-analysis.''

\subsection{Basic framework}
\begin{assumption}[BAMA model]
\label{ass:BAMA-model}
Consider \(J\) independent studies. In study \(j\) $(j=1, \ldots, J)$, let \(\hat\tau_j\) be the effect estimate with known variance \(\sigma_j^2\), and assume 
\[
    \hat\tau_j\sim N(\tau_j,\sigma_j^2)
\]
for an unknown study-specific effect \(\tau_j\). 
Fix an unknown target effect \(\tau\) and decompose $\tau_j$ as 
\[
    \tau_j=\tau+\Delta_j,
\]
where
\[
    \bm{\Delta}=(\Delta_1,\ldots,\Delta_J)^\top
\]
is the fixed, unknown bias vector. Fix a given bias level $b\geq0$. We assume $\bm{\Delta}\in\cD_b$, where
\[
    \cD_b
    =
    \left\{
        \bm{\Delta}\in\R^J:
        \left|
        \frac{\Delta_j}{\sqrt J\,\sigma_{\FE}}
        \right|
        \leq b
        \text{ for all }j=1,\ldots,J
    \right\}.
\]
\end{assumption}

In Assumption~\ref{ass:BAMA-model}, each study estimates its own effect \(\tau_j\), which can differ from the
target $\tau$, but the bias level $b$ controls the magnitude of this difference. We impose no distributional assumption on the \(\Delta_j\)'s, so the biases may vary arbitrarily across studies within the specified bound.

If $\bm{\Delta}$ were unrestricted, $\tau$ could be arbitrary. The restriction $\bm{\Delta}\in\cD_b$ in Assumption \ref{ass:BAMA-model} permits nontrivial statistical inference. The BAMA model with $b=0$ reduces to the FEMA model. Increasing $b$
allows progressively larger biases deviating from the FEMA model. Rather
than selecting a single value of $b$, we construct a sequence of confidence
intervals and examine how the resulting confidence interval for $\tau$ varies with $b$. In particular, this sequence allows us to assess how large the
study-specific biases must be before a statistically significant result becomes
insignificant.

The sensitivity parameter $b$ bounds each study's standardized bias, rather than its unstandardized bias $\Delta_j$, so $b$ is invariant to changes in measurement units. We standardize each bias by $\sqrt J\,\sigma_{\FE}$, where the factor $\sqrt J$ rescales the FEMA standard error $\sigma_{\FE}$ to the scale of a single study's standard error. To interpret this denominator, consider the homoskedastic case where all studies have the same standard error $\sigma_j = \sigma$. Then $\sqrt J\,\sigma_{\FE} = \sigma$, so each study's bias is bounded in absolute value by $b$ times a single-study standard error. When the standard errors differ, the denominator provides a common reference scale for all studies. This standardization parallels measures of effect size such as Cohen's $d$ and Glass's $\Delta$ \citep{cohen1988,glass1981}, which express effects in units of an outcome standard deviation.

\begin{remark}[Relation to other meta-analysis models]
\label{rem:fixed-effects-BAMA}
The fixed-effect\textit{s} model of \citet{rice2018} also treats the
study-specific effects $\tau_1,\ldots,\tau_J$ as fixed, unknown quantities.
Their biases relative to a fixed target $\tau$ are therefore fixed and
unknown, as in BAMA. Under their fixed-effect\textit{s} model,
$\hat\tau_{\FE}$ and its Wald-type confidence interval target the
precision-weighted average of the $\tau_j$'s.
\citet{rice2018} interpreted this
target as the estimand from a combined individual-level analysis that
adjusts for study. Thus, the FEMA inference can remain meaningful
under heterogeneous effects. 
This view appeared in \citet{fisher1935design}, which motivated \citet{yatescochran1938} to develop an early version of \citet{cochran1954}'s meta-analysis. Recently, \citet{qiu2025mantel} adopted a similar view, and \citet{berenfeld2025causal} discussed it from a causal inference perspective. 
Relatedly, \citet{claggett2014}  treated the study-specific effects as fixed, unknown parameters, but targeted quantiles of these fixed effects. We differ from these formulations and focus on $\tau$ defined in the BAMA model.
\end{remark}

\begin{remark}[Relation to other sensitivity analyses for meta-analysis] Our BAMA model allows the biases $\Delta_j$'s to be unstructured except for the upper bound $b$. Other sensitivity analysis methods for meta-analysis focus on specific forms of bias, e.g. confounding bias \citep{mathur2020unmeasured} and publication bias \citep{mathur2020publication}. It is of interest to extend our framework to their settings.
\end{remark}

\subsection{Confidence intervals and the maximum bias for significance}\label{sec::BAMA-inference}

Now we discuss statistical inference under the BAMA model. 
We will propose confidence intervals based on the point estimator and standard error based on the FEMA model, $\hat\tau_{\FE}$ and $\sigma_{\FE}$. Consider the symmetric interval
\begin{equation}
\label{eq:treatment-control-centered-interval}
    [\hat\tau_{\FE}-\sigma_{\FE}\cdot L,\,
    \hat\tau_{\FE}+\sigma_{\FE}\cdot L].
\end{equation}
We need to specify $L$ to ensure the correct coverage rate of $\tau$ under the BAMA model.
We retain the FEMA estimate as the center and modify only the radius,
which gives a direct extension of the Wald-type confidence interval under FEMA.
Without any bias under FEMA, the required radius is the standard-normal
quantile $z_{1-\zeta/2}$. Under Assumption~\ref{ass:BAMA-model}, the required radius is instead a folded-normal quantile. We will provide the formal results below. 

To present Theorem \ref{thm:treatment-control-ci} below, we need to introduce additional notation. For $x\geq 0$, let $\Phi(\cdot)$ denote the cumulative distribution function of the standard normal distribution
 and $L_\zeta(x)$ denote the $(1-\zeta)$ quantile of the folded-normal distribution $|N(x,1)|$.
Write
\[
    t_{\FE}=\frac{\hat\tau_{\FE}}{\sigma_{\FE}}
\]
for the $t$-statistic under FEMA.

\begin{theorem}[BAMA confidence interval and maximum bias for significance]
\label{thm:treatment-control-ci}
\label{thm:treatment-control-bvalue}
\label{thm:treatment-control-bvalue-asymp}
Fix $\zeta\in(0,1/2)$ and suppose Assumption~\ref{ass:BAMA-model} holds.
\begin{enumerate}[label=(\alph*)]
\item The shortest symmetric confidence interval in the form of
\eqref{eq:treatment-control-centered-interval} has 
\[ 
L=L_\zeta(\sqrt J \, b) .
\]
\item Let $\overline{b}(\zeta)$ be the smallest $b\geq0$ for which the confidence interval in (a)
contains zero.  If $|t_{\FE}|\leq z_{1-\zeta/2}$, then $\overline{b}(\zeta)=0$. Otherwise, the
following equation uniquely determines $\overline{b}(\zeta)>0$:
\[
    \Phi\left\{|t_{\FE}|-\sqrt J\cdot\overline{b}(\zeta)\right\}
    -\Phi\left\{-|t_{\FE}|-\sqrt J\cdot\overline{b}(\zeta)\right\}
    =1-\zeta.
\]
\item 
The $\overline{b}(\zeta)$ in (b) has a first-order approximation: 
\[
    \overline{b}(\zeta)
    -\frac{|t_{\FE}|}{\sqrt J}
    \xrightarrow{\PP}0,
\]
as $J\to\infty$. 
\end{enumerate}
\end{theorem}

By Theorem~\ref{thm:treatment-control-ci}(a), the critical value changes from a standard-normal quantile to a folded-normal quantile when there is bias under BAMA. The maximum bias of $t$-statistics is at most $\sqrt J\,b$ under BAMA,                  
so the least-favorable absolute error has distribution $|N(\sqrt J\,b,1)|$.
When $b=0$, this distribution reduces to $|N(0,1)|$, whose $(1-\zeta)$ quantile
is $z_{1-\zeta/2}$, and the interval in Theorem~\ref{thm:treatment-control-ci}(a) reduces to the Wald-type confidence interval under FEMA. The confidence interval is sharp in the sense that the confidence interval gives exact $(1-\zeta)$ coverage for some $\bm{\Delta}\in\mathcal{D}_b$. 
The same interval remains asymptotically valid without the normality assumption on the $\hat{\tau}_j$'s; see Appendix~\ref{app:BAMA-details}.

The BAMA $p$-value corresponding to
Theorem~\ref{thm:treatment-control-ci}(a), for testing $\tau=0$
at a fixed bias level $b$, is
\begin{equation}
\label{eq:BAMA-p-value}
    1-\Phi(|t_{\FE}|-\sqrt J\,b)
    +\Phi(-|t_{\FE}|-\sqrt J\,b).
\end{equation}
At $b=0$, it reduces to the $p$-value under FEMA.
As $b$ increases, it is nondecreasing, reflecting the
weaker evidence against $\tau=0$ when larger biases are allowed.
The confidence interval in Theorem~\ref{thm:treatment-control-ci}(a)
excludes zero if and only if the BAMA $p$-value is less than $\zeta$, which induces Theorem~\ref{thm:treatment-control-bvalue}(b).

The phase-transition value $\overline{b}(\zeta)$ in Theorem~\ref{thm:treatment-control-bvalue}(b) is related to the notion of the {\it b-value} in \citet[Definition~2.2]{lin2026}. We  refer to it as the {\it maximum bias for significance} under BAMA. It is also related to the {\it E-value} in sensitivity analysis for unmeasured confounding in causal inference with observational studies \citep{vanderweele2017evalue}.

The required radius in Theorem~\ref{thm:treatment-control-ci}(a) is the $(1-\zeta)$ quantile of $|N(\sqrt J\,b,1)|$ and increases with $b$. Thus, the BAMA confidence interval widens with $b$ and contains zero for every $b\geq\overline{b}(\zeta)$. When the FEMA interval excludes zero, the maximum bias for significance is the bias level at which the interval first touches zero. We can plot the confidence limits against $b$ and read $\overline{b}(\zeta)$ from their first intersection with zero.

By Theorem~\ref{thm:treatment-control-bvalue-asymp}(c), with a large number of studies, the maximum bias for significance is
approximately the magnitude of the FEMA $t$-statistic divided by $\sqrt J$. This approximation does not depend on $\zeta$. Obviously, its exact distribution depends on $\zeta$, but we omit the details. 

The FEMA reports the $p$-value using $t_{\FE}$ and the standard normal quantile. This $p$-value measures evidence against the null hypothesis of zero $\tau$ under the FEMA model. In contrast, we propose to focus on the FEMA $t$-statistic divided by $\sqrt J$, which is a first-order approximation of the maximum bias for significance under BAMA. This approximation in Theorem \ref{thm:treatment-control-bvalue-asymp}(c) is intuitive, because the null result arises when the standardized bias exceeds the effect size.

\section{Model checking of BAMA}
\label{sec:model-checking}

Like other statistical models, the BAMA model in Assumption~\ref{ass:BAMA-model} with a given bias level $b$ can be rejected by the observed data. 
In Section \ref{sec::model-check-test}, we will develop a formal statistical test for model checking of BAMA, which leads to the notion of the {\it minimum bias for compatibility}. 
In Section \ref{sec::compare-lowerb-upperb}, we will compare the minimum bias for compatibility with the maximum bias for significance introduced in Section \ref{sec:BAMA}. 
In Section \ref{sec::optimistic-ci}, we will discuss an optimistic
confidence interval and $p$-value, both evaluated
at the minimum bias for compatibility.

\subsection{A statistical test for BAMA and the minimum  bias for compatibility}\label{sec::model-check-test}

Under the FEMA model in Assumption~\ref{ass:FEMA-model}, Cochran's \(Q\) statistic follows
a \(\chi^2_{J-1}\) distribution, and this null distribution allows us to test for the presence of between-study heterogeneity. We extend this
test to the BAMA model in Assumption~\ref{ass:BAMA-model} at a given bias level
\(b\). Based on $Q$, the key modification is that we must use the noncentral \(\chi^2_{J-1}\) distribution under
the BAMA model.

Let $\operatorname{pchisq}(\cdot;\mathrm{df},\mathrm{ncp})$ denote the cumulative
distribution function of a noncentral $\chi^2$ variable with
$\mathrm{df}$ degrees of freedom and noncentrality parameter $\mathrm{ncp}$,
following the notation of R. Recall $\widehat\sigma_{\textup{B}}^2$ in \eqref{eq:hatsigmaB} as the estimator for the between-study variance under the REMA model.

\begin{theorem}[Model checking of BAMA based on $Q$]
\label{thm:treatment-control-Q}
\label{thm:BAMA-model-checking threshold-asymptotics}
Fix $\zeta\in(0,1/2)$, and suppose that $J\geq2$.
\begin{enumerate}[label=(\alph*)]
\item Under Assumption~\ref{ass:BAMA-model}, \[
    p_Q(b)
    =1-\operatorname{pchisq}
    (Q;\mathrm{df}=J-1,\mathrm{ncp}=\Lambda_b),
\]
with
\[
    \Lambda_b=Jb^2\left\{1-
    \min_{s_1,\ldots,s_J\in\{-1,1\}}
    \left(\sum_{j=1}^J\lambda_{j,\FE}s_j\right)^2\right\},
\] is a valid
$p$-value. 

\item Let \(\underline{b}(\zeta)\) be the smallest bias level $b$ such that
$p_Q(b)$ is larger than or equal to $\zeta$. As $J\to\infty$, suppose that
\[
    \max_{1\leq j\leq J}\lambda_{j,\FE}\longrightarrow0,
    \qquad
    \frac{\widehat\sigma_{\textup{B}}}{\sqrt J\,\sigma_{\FE}}
    =O_{\PP}(1).
\]
Then
\[
    \underline{b}(\zeta)
    -
    \frac{\widehat\sigma_{\textup{B}}}{\sqrt J\,\sigma_{\FE}}
    \xrightarrow{\PP}0.
\]
\end{enumerate}
\end{theorem}

By Theorem~\ref{thm:treatment-control-Q}(a), we can use $Q$ to test whether the BAMA model is rejected by the observed data. 
At $b=0$, $p_Q(b)$ reduces to the $p$-value of Cochran's $Q$-test under FEMA.
As $b$ increases, $p_Q(b)$ is nondecreasing.

The first condition in part (b) requires that no study dominates
the precision weights, which is equivalent to the study variances $\sigma_j^2$ being uniformly bounded above and away from zero. The second requires that the estimated between-study standard deviation $\widehat\sigma_{\textup{B}}$ divided by the common reference scale $\sqrt J\,\sigma_{\FE}$ is bounded in probability, which ensures the boundedness of the approximation for $\underline{b}(\zeta)$.

By the definition in Theorem~\ref{thm:treatment-control-Q}(b), $\underline{b}(\zeta)$ is the bias level at which $p_Q(b)$ first reaches $\zeta$. Hence, the \(Q\)-test at level \(\zeta\) rejects the BAMA model at every bias level $b<\underline{b}(\zeta)$. We call this threshold $\underline b(\zeta)$ the
{\it minimum bias for compatibility} under BAMA, because it is the smallest
bias level at which the $Q$-test at level $\zeta$ does not reject BAMA.
We can plot
$p_Q(b)$ against $b$ and read the minimum bias for compatibility from the
curve's first intersection with the horizontal line at $\zeta$ when Cochran's $Q$-test rejects the FEMA model.

Theorem~\ref{thm:BAMA-model-checking threshold-asymptotics}(b) shows that, with a large number of studies, the minimum bias for compatibility is approximately the estimated between-study standard deviation $\widehat\sigma_{\textup{B}}$ divided by the common reference scale $\sqrt J\,\sigma_{\FE}$. 
Theorem~\ref{thm:BAMA-model-checking threshold-asymptotics}(b) is related to REMA. 
Under REMA, $\widehat\sigma_{\textup{B}}$ estimates the between-study standard deviation.
The squared approximation, $\hat{\sigma}^2_{\textup{B}} / (J \sigma_{\FE}^2)$, is therefore approximately the ratio of the estimated between-study variance to the harmonic mean of the within-study variances. Thus, greater between-study variation relative to within-study variation requires a larger bias level for the model checking not to reject BAMA.

\begin{remark}[Computing the minimum bias for compatibility]
\label{rem:computation-basic}
Computing the $p$-value in Theorem \ref{thm:BAMA-model-checking threshold-asymptotics}(a) involves the minimization problem in $\Lambda_b$, which is a weighted two-way number-partitioning problem.
Exact sign enumeration requires exponential time.
For moderate $J$, a meet-in-the-middle algorithm reduces the running time to $O(J2^{J/2})$ and requires $O(2^{J/2})$ memory \citep{horowitz1974computing}.
For large $J$, we can replace $\Lambda_b$ by $Jb^2$ to obtain a
conservative $p$-value, which approximates the exact $p$-value under the no-dominant-study condition in Theorem~\ref{thm:BAMA-model-checking threshold-asymptotics}(b). 
\end{remark}

\subsection{Comparing the minimum bias for compatibility with the maximum bias for significance}
\label{sec::compare-lowerb-upperb}

Section \ref{sec::model-check-test} has introduced the notion of the minimum bias for compatibility, whereas Section \ref{sec::BAMA-inference} has introduced the notion of the maximum bias for significance. They quantify different aspects of the BAMA model. 
The minimum bias for compatibility is the model-checking threshold, below which the BAMA model is rejected by the observed data, whereas the maximum bias for significance is the significance threshold, above which we cannot reject the target null hypothesis under the BAMA model. 
Under Assumption~\ref{ass:BAMA-model}, they are independent because they are functions of $Q$ and $t_{\FE}$, respectively, which are independent.

If the minimum bias for compatibility is smaller than the maximum bias for significance, then the bias values in the interval $[\underline{b}(\zeta),\overline{b}(\zeta))$ are of particular interest. Those bias values are compatible with the observed data and, at the same time, yield significant results under the BAMA model. Outside this interval, the bias values are either incompatible with the observed data or yield insignificant results. Therefore, a key methodological advance of our paper is to recommend that practitioners report the interval 
\begin{equation}
\label{eq::interval-report}
[\underline{b}(\zeta),\overline{b}(\zeta)).
\end{equation}

However, it is perhaps surprising that the minimum bias for compatibility can be even larger than the maximum bias for significance, that is, the interval \eqref{eq::interval-report} is empty. If this happens, we declare a null result because there are no bias levels that are compatible with the data and, at the same time, result in significant results under the BAMA model. It is therefore interesting to investigate this regime in more detail below.

The asymptotic analyses in Theorems  \ref{thm:BAMA-model-checking threshold-asymptotics}(b) and \ref{thm:treatment-control-bvalue-asymp}(c) demonstrate that under BAMA, we have 
\begin{equation}
\label{eq::compare-bs}
\underline{b}(\zeta) \approx \frac{\widehat\sigma_{\textup{B}}}{\sqrt J\,\sigma_{\FE}},\quad 
\overline{b}(\zeta) \approx  \frac{|\hat\tau_{\FE}|}{\sqrt J\,\sigma_{\FE}} ,
\end{equation}
with the key quantities being the number of studies $J$, the point estimate and standard error under FEMA, and the estimator of the between-study standard error under REMA.  Therefore, 
the magnitude of the minimum bias for compatibility relative to the maximum bias for significance depends on that of the FEMA estimate relative to the estimated between-study standard deviation. 

To gain more insight, we will compare them under the FEMA and REMA models below.

\begin{proposition}
    \label{prop::compare-bs}
    Suppose that $\tau$ and $\sigma_{\textup{B}}^2$ are fixed, and that all studies have the same positive variance $\sigma^2$.
Then the denominator of both first-order approximations reduces to $\sigma$. 
\begin{enumerate}[label=(\alph*)]
\item 
Under FEMA, we have 
\[
    \underline{b}(\zeta)\xrightarrow{\PP}0,
    \qquad
    \overline{b}(\zeta)\xrightarrow{\PP}\frac{|\tau|}{\sigma}.
\]
\item 
Under REMA, we have 
\[
    \underline{b}(\zeta)\xrightarrow{\PP}\frac{\sigma_{\textup{B}}}{\sigma},
    \qquad
    \overline{b}(\zeta)\xrightarrow{\PP}\frac{|\tau|}{\sigma}.
\]
\end{enumerate}
\end{proposition}

Under equal study variances $\sigma^2$, the denominator in approximation \eqref{eq::compare-bs} reduces to $\sigma$. Proposition~\ref{prop::compare-bs} then follows from the consistency of the point estimate under FEMA and the between-study variance estimate under REMA.

From Proposition \ref{prop::compare-bs}(a), if the true model is FEMA, then the interval $[\underline{b}(\zeta),\overline{b}(\zeta))$ is nonempty with probability tending to one when $\tau\ne0$.
From Proposition \ref{prop::compare-bs}(b), if the true model is REMA, the interval $[\underline{b}(\zeta),\overline{b}(\zeta))$ is nonempty with probability tending to one when $|\tau|>\sigma_{\textup{B}}$ and empty with probability tending to one when $|\tau|<\sigma_{\textup{B}}$.

The following theorem makes a further connection between an empty $[\underline b(\zeta),\overline b(\zeta))$ and the traditional meta-analyses. 

\begin{theorem}
    \label{thm::empty-interval}
    Under Assumption \ref{ass:BAMA-model} with a fixed bias level $b\geq 0$ and $\zeta \in (0, 1/2)$, the interval $[\underline b(\zeta),\overline b(\zeta))$ is asymptotically empty if 
    \begin{enumerate}[label=(\alph*)]
\item 
we reject the FEMA model by the $Q$-test, and
\item 
we fail to reject the null value of $\tau$ based on $\hat\tau_{\FE}$ and $\hat{\sigma}_{\FE\dagger}^2$ discussed in Remark \ref{rem:FEMA-under-REMA}. 
    \end{enumerate}
\end{theorem}

By Theorem \ref{thm::empty-interval}, a null result under BAMA arises when the traditional FEMA model is rejected and the inference based on $\hat\tau_{\FE}$ and $\hat{\sigma}_{\FE\dagger}^2$ under the REMA model is null. Therefore, the traditional meta-analyses and our new meta-analysis largely agree on null results. They differ, however, for non-null results.

\subsection{An optimistic confidence interval and \texorpdfstring{$p$}{p}-value}
\label{sec::optimistic-ci}

We end this section with discussions of an optimistic confidence interval and $p$-value at the minimum bias for compatibility. The optimistic confidence interval is the confidence interval of \eqref{eq:treatment-control-centered-interval} with
$$
L = L_{\zeta} (\sqrt{J}\cdot \underline b(\zeta)). 
$$
It is the narrowest BAMA confidence interval among bias levels not rejected by the $Q$-test. The optimistic confidence interval need not be a valid confidence interval because nonrejection
by the $Q$-test does not imply that the BAMA model is true. Nevertheless, if it covers the null value of $\tau$, then the interval \eqref{eq::interval-report} must be empty. 

The optimistic $p$-value corresponding to the optimistic confidence interval is the $p$-value of \eqref{eq:BAMA-p-value} with $b=\underline{b}(\zeta)$. Similar to the optimistic confidence interval, it is the smallest BAMA $p$-value among bias levels not rejected by the $Q$-test and need not be a valid $p$-value. As in standard inference, the optimistic BAMA $p$-value is less than $\zeta$ if and only if the optimistic confidence interval excludes zero. We state the result below. 

\begin{proposition}
\label{prop::optimistic-ci-empty}
The following statements are equivalent:
\begin{enumerate}[label=(\alph*)]
\item $[\underline b(\zeta),\overline b(\zeta))$ is empty;
\item the optimistic confidence interval contains zero;
\item the optimistic $p$-value is greater than or equal to $\zeta$.
\end{enumerate}
\end{proposition}

Proposition~\ref{prop::optimistic-ci-empty} follows directly from the definitions. The optimistic confidence interval contains zero if and only if
$|t_{\FE}|\leq L_\zeta ( \sqrt J\cdot\underline b(\zeta) )$,
which is equivalent to
$\underline b(\zeta)\geq\overline b(\zeta)$ by monotonicity and the definition of $\overline b(\zeta)$. The equivalence with (c) follows from \eqref{eq:BAMA-p-value}.

The optimistic confidence interval need not have nominal coverage, and the
problem goes beyond using the data twice. The minimum bias for
compatibility provides a lower confidence limit on the bias level,
whereas the BAMA coverage guarantee requires an upper bound.
More generally, without restrictions on the biases, the study
estimates alone cannot distinguish a common bias from a shift in
the target effect. Splitting the significance level does not
resolve this identification problem and therefore does not
restore the coverage guarantee.

\section{Practical recommendations}
\label{sec:recommendations}

\subsection{What to report for meta-analysis}
\label{sec::what-to-report}

Although our paper is of a theoretical nature, we do want to provide recommendations for practitioners.

First, we recommend reporting the following basic statistics based on the classic FEMA and REMA:
\begin{itemize}
    \item 
    the number of studies $J$;
    \item 
    the point estimate $\hat\tau_{\FE}$, standard error $\sigma_{\FE}$, $t$-statistic $t_{\FE}= \hat\tau_{\FE} / \sigma_{\FE}$, confidence interval and $p$-value under FEMA, as well as Cochran's $Q$ statistic, its degrees of freedom and $p$-value for testing the FEMA model;
    \item 
    the point estimate $\hat\tau_{\RE}$, standard error $\sigma_{\RE}$, $t$-statistic $t_{\RE}= \hat\tau_{\RE} / \sigma_{\RE}$, confidence interval and $p$-value, as well as the estimate of the between-study standard error $\hat\sigma_{\textup{B}}$ under REMA;
    \item 
    the point estimate $\hat\tau_{\FE}$, standard error $\hat{\sigma}_{\FE\dagger}$, $t$-statistic $t_{\FE\dagger} = \hat\tau_{\FE} / \hat{\sigma}_{\FE\dagger}$, confidence interval and $p$-value under REMA, as discussed in Remark \ref{rem:FEMA-under-REMA}.
\end{itemize}

Second, we recommend reporting the following basic statistics based on the proposed BAMA:
\begin{itemize}
    \item 
for a sequence of $b$,  the confidence intervals and $p$-values based on the FEMA point estimate $\hat\tau_{\FE}$ and standard error $\sigma_{\FE}$, based on the quantiles of the folded-normal distribution $|N(\sqrt J\,b,1)|$, according to Theorem \ref{thm:treatment-control-ci}(a);
\item 
the maximum bias for significance $\overline{b}(\zeta)$ and its first-order approximation $|t_{\FE}| / \sqrt J$, according to Theorem \ref{thm:treatment-control-ci}(b) and (c);
\item 
for a sequence of $b$, the model-checking $p$-value $p_Q(b)$, according to Theorem \ref{thm:treatment-control-Q}(a);
\item 
the minimum bias for compatibility $\underline{b}(\zeta)$ and its first-order approximation $\widehat\sigma_{\textup{B}} / (\sqrt J\,\sigma_{\FE})$, according to Theorem \ref{thm:treatment-control-Q}(b).
\end{itemize}

Third, we recommend checking whether BAMA yields a null result using the following three equivalent conditions
(see Proposition~\ref{prop::optimistic-ci-empty}):
\begin{itemize}
    \item 
    whether the interval $[\underline{b}(\zeta),\overline{b}(\zeta))$ is empty;
    \item 
    whether the optimistic confidence interval in Section \ref{sec::optimistic-ci} includes the null value.
    \item whether the optimistic $p$-value in Section \ref{sec::optimistic-ci} is larger than or equal to $\zeta$.
\end{itemize}

We will illustrate the above recommendations in Section \ref{sec:examples} below.

\subsection{Calibrating the maximum bias for significance}
\label{sec::calibrating-b-value}

The BAMA model in Assumption \ref{ass:BAMA-model} requires specifying the bias parameter $b$, whose upper bound is beyond the data without further assumptions. 
Based on Section \ref{sec::what-to-report}, if BAMA results in a null result, we do not need to further reason about the plausible upper bound of the bias parameter $b$. However, if the optimistic confidence interval in Section \ref{sec::optimistic-ci}  excludes the null value, we need to calibrate the maximum bias for significance $\overline{b}(\zeta)$.

Since the information on the upper bound of $b$ is beyond the data without additional assumptions, we must calibrate $\overline{b}(\zeta)$ based on domain knowledge, e.g., based on the bias of previous, related studies. If $\overline{b}(\zeta)$ is larger than the reasonable range of the bias, BAMA results in a significant result despite the potential study-specific bias. Otherwise, BAMA results in a null result.

A practical approach to calibrate $\overline{b}(\zeta)$ is leave-one-study-out, that is, calculating the empirical bias due to omitting one study in the data. Equivalently, we can plot the histogram of the standardized
absolute differences 
$$
b_j =  \frac{ |\hat\tau_{\FE}-\hat\tau_j|}{\sqrt J\,\sigma_{\FE}}, \quad (j=1, \ldots, J)
$$ 
to obtain the range of the bias. We will illustrate this approach in Section \ref{sec:examples} below. 

\section{Example}
\label{sec:examples}

We illustrate our method by reanalyzing the depression treatments comparison of \citet{williams2026}.
We set $\zeta=0.05$ throughout.

\label{sec:williams-treatment-control-example}

\citet{williams2026} synthesized changes from baseline in the Hamilton
Depression Rating Scale, where negative values indicate improvement.
The source analysis contained 16 open-label traditional-antidepressant (TAD)
trials and 8 psychedelic-assisted-therapy (PAT) trials.  Figure~2A of their article
displays model-based posterior medians and $95\%$ credible
intervals for each study.  To keep the trial as the independent unit
without imposing an additional covariance model, we exclude the three-arm Arns
et al.\ (2016) trial.  The resulting reanalysis contains 15 TAD trials and 8 PAT
trials.

We analyze the TAD and PAT studies separately.  To echo Section \ref{sec:recommendations}, Table~\ref{tab:williams-nmasens-results} summarizes the FEMA and REMA results in Panel A and the BAMA results in Panel B.

First, Panel A
of Table~\ref{tab:williams-nmasens-results} reports the basic statistics based on FEMA and REMA. The estimated between-study standard deviations are large compared with the FEMA standard errors. Consequently, the REMA standard errors and the REMA-adjusted standard errors $\hat{\sigma}_{\FE\dagger}$ are substantially larger than the FEMA standard errors. All three confidence intervals exclude zero
for both treatments. Cochran's $Q$-test nevertheless rejects the
FEMA model in both analyses. 

\begin{table}[htbp]
\centering
\small
\caption{Reanalysis of \cite{williams2026}. The subscript $_\dagger$ denotes the inference based on the FEMA point estimate and the inflated variance estimator in Remark~\ref{rem:FEMA-under-REMA}.}
\label{tab:williams-nmasens-results}
{\setlength{\tabcolsep}{4pt}
\begin{tabular*}{\textwidth}{@{\extracolsep{\fill}}llrrrrr@{}}\\[0em]
\multicolumn{7}{l}{\textbf{Panel A. FEMA and REMA}}\\[0em]
\toprule
Treatment & Procedure & Estimate & SE & $t$-statistic & 95\% Confidence interval & $p$-value \\
\midrule
\multirow{3}{*}{Open-label TADs}
 & FEMA     & $-13.078$ & $0.162$ & $-80.944$ & $[-13.395,-12.762]$ & $<.001$ \\
 & REMA     & $-12.603$ & $0.680$ & $-18.538$ & $[-13.935,-11.270]$ & $<.001$ \\
 & FEMA$_\dagger$ & $-13.078$ & $0.855$ & $-15.298$ & $[-14.754,-11.403]$ & $<.001$ \\
\addlinespace
\multirow{3}{*}{PAT}
 & FEMA     & $-11.267$ & $0.637$ & $-17.686$ & $[-12.515,-10.018]$ & $<.001$ \\
 & REMA     & $-11.530$ & $0.988$ & $-11.666$ & $[-13.467,-9.593]$  & $<.001$ \\
 & FEMA$_\dagger$ & $-11.267$ & $1.039$ & $-10.841$ & $[-13.303,-9.230]$  & $<.001$ \\
\bottomrule
\end{tabular*}}

\par\smallskip
\noindent\footnotesize Open-label TADs: $J=15$, $Q=228.409$ on 14 degrees of freedom, $p_Q(0)<0.001$, $\widehat\sigma_{\textup{B}}^2=6.302$;\par PAT: $J=8$, $Q=15.215$ on 7 degrees of freedom, $p_Q(0)=0.033$, $\widehat\sigma_{\textup{B}}^2=4.008$.

\vspace{0.75em}
{\setlength{\tabcolsep}{4pt}
\begin{tabular*}{\textwidth}{@{\extracolsep{\fill}}lrrrrrr@{}}
\multicolumn{7}{l}{\textbf{Panel B. BAMA}}\\[0em]
\toprule
& \multicolumn{2}{c}{Optimistic} & \multicolumn{2}{c}{maximum bias for significance} & \multicolumn{2}{c}{minimum bias for compatibility} \\
\cmidrule(lr){2-3}\cmidrule(lr){4-5}\cmidrule(l){6-7}
Treatment & Confidence interval & $p$-value & Exact & Approximation & Exact & Approximation \\
\midrule
Open-label TADs & $[-15.445,-10.711]$ & $<.001$ & $20.475$ & $20.900$ & $3.358$ & $4.012$ \\
PAT             & $[-12.806,-9.727]$  & $<.001$ & $5.671$  & $6.253$  & $0.270$ & $1.111$ \\
\bottomrule
\end{tabular*}}
\end{table}

\begin{figure}[htbp]
\centering
\begin{tikzpicture}
\node[inner sep=0, outer sep=0] (fig) {
    \includegraphics[width=0.9\textwidth]
    {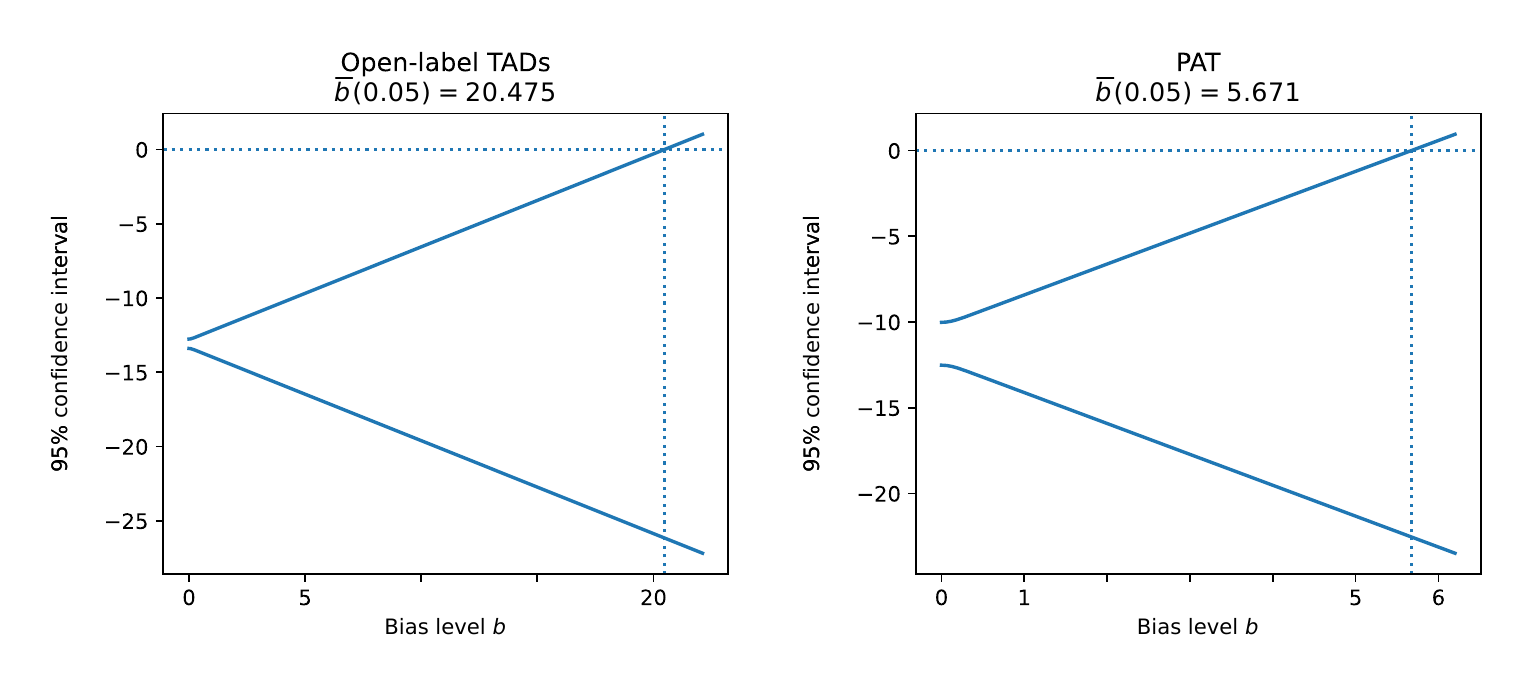}
};
\node[anchor=north west, xshift=4pt, yshift=-4pt]
    at (fig.north west) {\textbf{(a)}};
\node[anchor=north west, xshift=4pt, yshift=-4pt]
    at (fig.north) {\textbf{(b)}};
\end{tikzpicture}
\caption{BAMA confidence intervals from the reconstructed trial-level summaries in
\citet{williams2026}. The left and right panels show the TAD and PAT analyses, respectively.
For each treatment, the solid curves are the lower and upper
endpoints, the horizontal line marks zero change, and the vertical dotted line
marks the maximum bias for significance.}
\label{fig:williams-treatment-control-plot}
\end{figure}

\begin{figure}[htbp]
\centering
\includegraphics[width=0.95\textwidth]{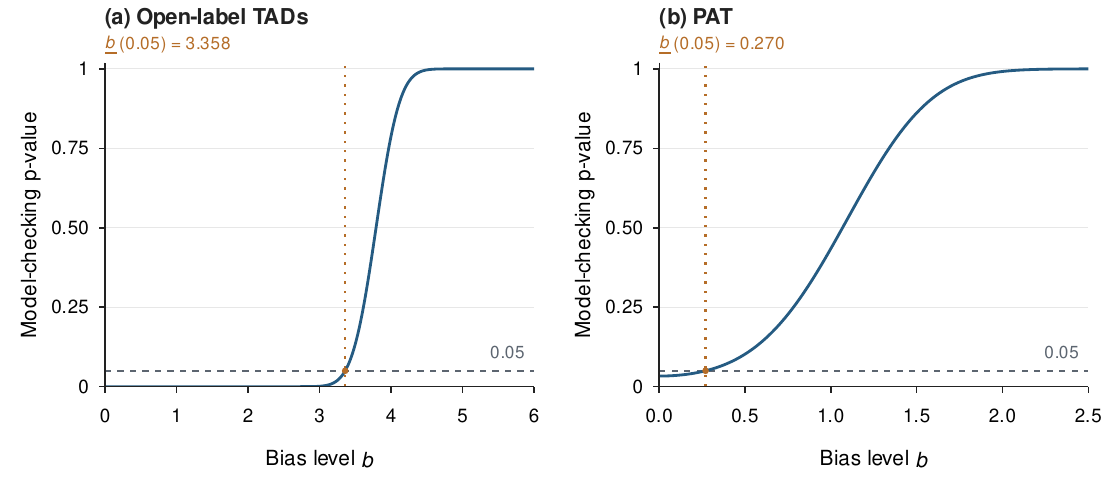}
\caption{Model checking of BAMA using the reconstructed trial-level
summaries in \citet{williams2026}. Panels (a) and (b) show the
open-label TAD and PAT analyses, respectively. The solid curves show
$p_Q(b)$, the horizontal dashed line marks $0.05$, and the vertical
dotted lines mark the minimum bias for compatibility. The panels
use different horizontal scales.}
\label{fig:williams-model-checking}
\end{figure}

Second, Panel~B of Table~\ref{tab:williams-nmasens-results}, Figure~\ref{fig:williams-treatment-control-plot} and Figure~\ref{fig:williams-model-checking} report the BAMA results. Figure~\ref{fig:williams-treatment-control-plot} displays the BAMA confidence intervals at different bias levels $b$ and the maximum bias for significance. At $b=0$, they reduce to the FEMA confidence intervals, which exclude zero. As $b$ increases, they widen and first include zero at the maximum bias for significance. Figure~\ref{fig:williams-model-checking} displays the model-checking
$p$-values at different bias levels. The minimum bias for
compatibility is the bias level at which each curve first reaches
$0.05$.

For both treatments, the minimum bias for compatibility is positive,
so some study-specific bias must be allowed for the $Q$-test not to
reject BAMA. The maximum bias for significance is larger than the
minimum bias for compatibility. Therefore, the bias levels in
$[\underline b(0.05),\overline b(0.05))$ are not rejected by the
model check and yield confidence intervals that exclude zero.
The optimistic confidence intervals also exclude zero and the optimistic $p$-values are less than $0.05$, in agreement with
Proposition~\ref{prop::optimistic-ci-empty}.

The large maximum bias levels for significance for both treatments indicate that substantial study-specific bias can be allowed before the BAMA confidence intervals
include zero. The minimum bias levels for compatibility are much smaller, so the observed heterogeneity can be accommodated with a smaller bias level than is needed for the results to lose statistical significance. Practitioners can
assess whether these bias levels are reasonable based on domain knowledge. 

\begin{figure}[htbp]
\centering
\includegraphics[width=0.95\textwidth]{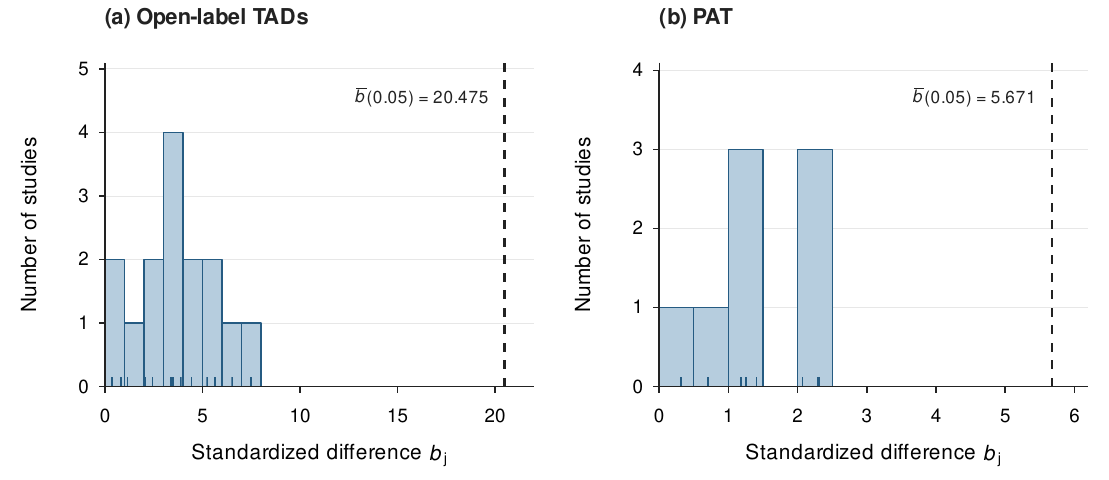}
\caption{Empirical calibration of the maximum bias for significance.
Panels (a) and (b) show histograms of
$b_j=|\hat\tau_{\FE}-\hat\tau_j|/(\sqrt J\,\sigma_{\FE})$
for the open-label TAD and PAT analyses, respectively.
The vertical dashed lines mark the maximum bias for significance,
and the short marks along the horizontal axes show the individual
$b_j$ values. The panels use different horizontal scales.
}
\label{fig:williams-calibration}
\end{figure}

Because the optimistic confidence intervals exclude zero, we further consider
calibration of the maximum bias for significance, following
Section~\ref{sec::calibrating-b-value}. Figure~\ref{fig:williams-calibration} shows that all standardized
differences $b_j$ lie below the corresponding maximum bias for
significance. The standardized differences $b_j$ provide an empirical reference for
the bias level in each analysis, although they cannot detect a bias
shared by all studies. 

\section{Conclusion and future work}
\label{sec:discussion}

This paper proposes BAMA that treats
study-level biases as fixed but bounded unknown quantities. It departs from the
classic FEMA and REMA formulations in that it neither assumes a common parameter across studies nor
assumes the study-specific biases to be mean-zero random variables. 
Under BAMA, we have discussed statistical inference and highlighted two important notions: the maximum bias for significance that equals the threshold beyond which the bias explains away the statistical results, and the minimum bias for compatibility that equals the threshold below which the bias is too small for the model to be compatible with the observed data. 

We have developed BAMA for treatment-control studies. We will communicate the corresponding results for network meta-analysis in a separate report. 

\section*{Data availability}
The data analyzed in this paper are from the publicly available Figure~2A of \citet{williams2026}. We implement our method in the R package \texttt{BAmeta}, which is available together with the replication materials at \url{https://github.com/mingxuanzou/BAmeta}.

\section*{Acknowledgments}
Abdullah Ateyeh gave us helpful comments. Peng Ding is partially supported by the U.S. National Science Foundation (\# 2514234). We use ChatGPT and Claude to polish the text.

\bibliographystyle{abbrvnat}
\bibliography{BAMA_references}

\clearpage
\setcounter{page}{1}
\renewcommand{\thepage}{S\arabic{page}}
\appendix
\numberwithin{equation}{section}

\setcounter{assumption}{0}
\setcounter{definition}{0}
\setcounter{theorem}{0}
\setcounter{lemma}{0}
\setcounter{remark}{0}
\renewcommand{\theassumption}{\thesection.\arabic{assumption}}
\renewcommand{\thedefinition}{\thesection.\arabic{definition}}
\renewcommand{\thetheorem}{\thesection.\arabic{theorem}}
\renewcommand{\thelemma}{\thesection.\arabic{lemma}}
\renewcommand{\theremark}{\thesection.\arabic{remark}}
\setcounter{figure}{0}
\setcounter{table}{0}
\renewcommand{\thefigure}{\thesection.\arabic{figure}}
\renewcommand{\thetable}{\thesection.\arabic{table}}

\begin{center}
{\Large\bfseries Supplementary Materials}
\end{center}

Appendix~\ref{app:additional-results} states the additional results that complement the main text.
\begin{enumerate}[label=(\alph*)]
    \item Appendix~\ref{app:standard-relaxations} establishes the
    large-sample results for the REMA estimator and the FEMA estimator with inflated variance in Section~\ref{sec:background}, without assuming normality.
    \item Appendix~\ref{app:BAMA-details} establishes asymptotic validity of the BAMA confidence interval in Section~\ref{sec:BAMA}, without assuming normality.
\end{enumerate}

Appendix~\ref{app:proofs} contains the auxiliary lemmas and all proofs.
\begin{enumerate}[label=(\alph*)]
    \item Appendix~\ref{app:auxiliary-lemmas} states and proves lemmas on folded-normal and noncentral chi-square inversion.
    \item Appendix~\ref{app:proofs-REMA} proves the large-sample results for the REMA estimator and the FEMA estimator with inflated variance in Section~\ref{sec:background} and Appendix~\ref{app:standard-relaxations}.
    \item Appendix~\ref{app:proofs-treatment-control} proves the BAMA results in Sections~\ref{sec:BAMA} and~\ref{sec:model-checking} and Appendix~\ref{app:BAMA-details}, including asymptotic validity and threshold ordering.
\end{enumerate}

\section{Additional results}
\label{app:additional-results}

\setcounter{assumption}{0}
\setcounter{definition}{0}
\setcounter{theorem}{0}
\setcounter{lemma}{0}
\setcounter{remark}{0}
\setcounter{proposition}{0}
\renewcommand{\theproposition}{\thesection.\arabic{proposition}}

\subsection{Additional results under REMA}
\label{app:standard-relaxations}

Section~\ref{app:standard-relaxations} justifies the REMA inference procedures that we review in
Section~\ref{sec:background} without assuming normality of the study
estimates or random effects (see Remark~\ref{rem:FEMA-under-REMA} in the main paper).

The following assumption relaxes Assumption~\ref{ass:REMA-model} by replacing
conditional normality of $\hat\tau_j\mid\tau_j$ with conditions on its first
two moments, and collects the regularity conditions used throughout this subsection.

\begin{assumption}[REMA model without normality and regularity conditions]
\label{ass:REMA-without-normality}
\label{ass:dl-regularity}
Retain the notation and independence of studies from
Assumption~\ref{ass:REMA-model}. For the study-specific effects
\(\tau_j\), suppose
\[
    \E(\hat\tau_j\mid\tau_j)=\tau_j,
    \qquad
    \Var(\hat\tau_j\mid\tau_j)=\sigma_j^2,
\]
and
\[
    \E(\tau_j)=\tau,
    \qquad
    \Var(\tau_j)=\sigma_{\textup{B}}^2,
\]
with unknown \(\sigma_{\textup{B}}^2\). Moreover, there exist universal constants $c_\sigma,M>0$ such that, for every $j$,
\[
    \E|\hat\tau_j-\tau|^{4}
    \leq M,
    \qquad
    \sigma_j\geq c_\sigma.
\]
\end{assumption}
By the laws of total expectation and total variance, this model gives
$\E(\hat\tau_j)=\tau$ and
$\Var(\hat\tau_j)=\sigma_j^2+\sigma_{\textup{B}}^2$.

\subsubsection{Asymptotic properties of REMA}
\label{app:dl-REMA-estimator}

To justify inference with the DerSimonian--Laird estimator, we need to
estimate $\sigma_{\textup{B}}^2$ consistently and control the error of the plug-in estimator. Assumption~\ref{ass:REMA-without-normality} supports both steps and the central limit
theorems below.

In Assumption~\ref{ass:REMA-without-normality}, the fourth-moment bound controls fluctuations of $Q$ and
the tails of the study estimates, while the lower bound on $\sigma_j$
prevents unbounded within-study precision.

The following theorem gives the convergence rate of the between-study variance estimator and the resulting central limit theorem.

\begin{theorem}[Central limit theorem for the REMA estimator]
\label{thm:dl-REMA-mean-clt}
Under Assumption~\ref{ass:REMA-without-normality}, as $J\to\infty$,
\begin{enumerate}[label=(\alph*)]
\item
\[
    \hat\sigma_{\textup{B}}^2-\sigma_{\textup{B}}^2=O_{\PP}(J^{-1/2}).
\]
\item
\[
    \frac{\hat\tau_{\RE}-\tau}{\hat\sigma_{\RE}}
    \Rightarrow N(0,1).
\]
\end{enumerate}
\end{theorem}

By Theorem~\ref{thm:dl-REMA-mean-clt}(b), the Wald-type confidence interval centered at
$\hat\tau_{\RE}$ with plug-in standard error $\hat\sigma_{\RE}$ is
asymptotically valid for $\tau$.

\subsubsection{FEMA estimator under REMA}

This subsubsection discusses the FEMA estimator under the REMA model and justifies Remark~\ref{rem:FEMA-under-REMA}. Under REMA, the
variance of the FEMA estimator must account for both sampling error and
between-study variation. 

\begin{theorem}[FEMA estimator under the REMA model]
\label{thm:FEMA-under-REMA}
Under Assumption~\ref{ass:REMA-without-normality},
\[
    \frac{\hat\tau_{\FE}-\tau}{\hat{\sigma}_{\FE\dagger}}
    \Rightarrow N(0,1).
\]
\end{theorem}

By Theorem~\ref{thm:FEMA-under-REMA}, the Wald-type confidence interval centered at $\hat\tau_{\FE}$ with standard error
$\hat{\sigma}_{\FE\dagger}$ is asymptotically valid for
$\tau$.

\subsection{Additional results under BAMA}
\label{app:BAMA-details}

Section~\ref{app:BAMA-details} provides an asymptotic version of BAMA without assuming
normality of the point estimates.

Under Assumption~\ref{ass:BAMA-model}, the unknown study-specific biases
enter the FEMA estimator through their precision-weighted average. Define this unknown bias relative to the target $\tau$ by
\[
    \Delta_{\FE}=\E(\hat\tau_{\FE})-\tau
    =
    \sum_{j=1}^J\lambda_{j,\FE}\Delta_j
    =
    \frac{\sum_{j=1}^J\sigma_j^{-2}\Delta_j}
    {\sum_{j=1}^J\sigma_j^{-2}}.
\]

We will show that the calculation in the proof of Theorem~\ref{thm:treatment-control-ci} also gives an asymptotic confidence interval under BAMA without the normality assumption.

\begin{proposition}[Asymptotic validity without normality]
\label{prop:treatment-control-nongaussian}
Fix $b\geq0$ and $\zeta\in(0,1/2)$. For each $J$ and
$\bm{\Delta}\in\cD_b$, suppose that the studies are independent and
satisfy
\[
    \E(\hat\tau_j)=\tau+\Delta_j,
    \qquad
    \Var(\hat\tau_j)=\sigma_j^2,
    \qquad j=1,\ldots,J,
\]
where the known variances $\sigma_j^2$ do not depend on $\bm{\Delta}$.
Suppose that there exist constants $c_\sigma,M_b>0$, independent of $J$,
such that
\[
    \sup_{\bm{\Delta}\in\cD_b}\max_{1\leq j\leq J}
    \E|\hat\tau_j-\tau-\Delta_j|^4
    \leq M_b,
    \qquad
    \min_{1\leq j\leq J}\sigma_j\geq c_\sigma.
\]
Then, as $J\to\infty$, the confidence interval in
Theorem~\ref{thm:treatment-control-ci} satisfies
\[
    \inf_{\bm{\Delta}\in\cD_b}
    \PP\left\{
        |\hat\tau_{\FE}-\tau|
        \leq\sigma_{\FE}L_\zeta(\sqrt J\,b)
    \right\}
    =1-\zeta+o(1).
\]
\end{proposition}

Proposition~\ref{prop:treatment-control-nongaussian} shows that for each fixed $b$, the confidence interval in Theorem~\ref{thm:treatment-control-ci} is asymptotically
valid and asymptotically sharp over $\cD_b$.

\section{Proofs}
\label{app:proofs}

\setcounter{lemma}{0}

\subsection{Lemmas}
\label{app:auxiliary-lemmas}

Lemma~\ref{lem:hzeta-large-signal} bounds the 
inverse folded-normal quantile, which we use to
derive the large-sample approximation to the maximum bias for significance.

\begin{lemma}
\label{lem:hzeta-large-signal}
Fix \(\zeta\in(0,1/2)\), and define\[
h_\zeta(s) 
=\inf\left\{
x\geq0:
L_\zeta(x)\geq s
\right\},
\qquad s\geq0.
\]
Then, for every \(s\geq0\),
\[
    0\leq s-h_\zeta(s)\leq z_{1-\zeta/2}.
\]
\end{lemma}

\begin{proof}[Proof of Lemma~\ref{lem:hzeta-large-signal}]
We first bound the difference between the folded-normal radius and its
noncentrality parameter, then invert that bound. For $x\geq0$, the
definition of $L_\zeta(x)$ gives
\[
    [1-\Phi\{L_\zeta(x)-x\}]+\Phi\{-L_\zeta(x)-x\}=\zeta.
\]
Since \(x\geq0\),
\[
    \Phi\{-L_\zeta(x)-x\}\leq \Phi\{x-L_\zeta(x)\}=1-\Phi\{L_\zeta(x)-x\}.
\]
It follows that
\[
    \frac{\zeta}{2}
    \leq
    1-\Phi\{L_\zeta(x)-x\}
    \leq
    \zeta,
\]
and hence
\[
    z_{1-\zeta}
    \leq
    L_\zeta(x)-x
    \leq
    z_{1-\zeta/2}.
\]
If \(0\leq s\leq L_\zeta(0)=z_{1-\zeta/2}\), then
\(h_\zeta(s)=0\), so
\[
    0\leq s-h_\zeta(s)\leq z_{1-\zeta/2}.
\]
If \(s>L_\zeta(0)\), let \(x=h_\zeta(s)\). Then
\(s=L_\zeta(x)\), and the preceding display gives the same bound.
\end{proof}

Lemma~\ref{lem:nchisq-uniform-inversion} approximates
the bias level obtained by inverting a noncentral chi-square tail
probability, which we use to derive the large-sample approximation to
the minimum bias for compatibility.

\begin{lemma}
\label{lem:nchisq-uniform-inversion}
Fix \(\zeta\in(0,1/2)\), and define
\[
h_{\nu,\zeta}(x)
=
\inf\left\{
b\geq0:
1-\operatorname{pchisq}
\left(
x;\mathrm{df}=\nu,\mathrm{ncp}=\nu b^2
\right)
\geq\zeta
\right\},
\qquad x\geq0.
\]
Then, as \(\nu\to\infty\),
\[
\sup_{x\geq0}
\left|
h_{\nu,\zeta}(x)
-
\sqrt{\frac{(x-\nu)_+}{\nu}}
\right|
\longrightarrow0.
\]
\end{lemma}

\begin{proof}[Proof of Lemma~\ref{lem:nchisq-uniform-inversion}]
We bound noncentral chi-square quantiles around their means and then
translate this bound into a uniform error for the inverse.

Let
\[
q_{1-\zeta}(\nu,\lambda)
=
\inf\left\{
x:
\operatorname{pchisq}
\left(
x;\mathrm{df}=\nu,\mathrm{ncp}=\lambda
\right)
\geq1-\zeta
\right\}.
\]
The mean and variance of a noncentral chi-square variable are
\(\nu+\lambda\) and \(2(\nu+2\lambda)\), respectively. Hence,
Chebyshev's inequality implies that
\[
\left|
q_{1-\zeta}(\nu,\lambda)-(\nu+\lambda)
\right|
\leq
C_\zeta\sqrt{\nu+2\lambda}
\]
for every $\nu\geq1$ and $\lambda\geq0$, with
$C_\zeta=\sqrt{2/\zeta}$.

Fix $x\geq0$. If \(h_{\nu,\zeta}(x)=0\), then
\(x\leq q_{1-\zeta}(\nu,0)\) and therefore
\[
\sqrt{\frac{(x-\nu)_+}{\nu}}\leq \frac{\sqrt{C_\zeta}}{\sqrt[4]{\nu}}\to 0.
\]

If \(h_{\nu,\zeta}(x)>0\), continuity and monotonicity in the noncentrality
parameter give
\[
x=q_{1-\zeta}\{\nu,\nu h_{\nu,\zeta}(x)^2\},
\]
and consequently
\[
\left|
\frac{x-\nu}{\nu}-h_{\nu,\zeta}(x)^2
\right|
\leq
\frac{C_\zeta}{\sqrt{\nu}}
\sqrt{1+2h_{\nu,\zeta}(x)^2}.
\]
If $h_{\nu,\zeta}(x)\leq1$, the preceding inequality gives
\[
\begin{aligned}
    \left|\sqrt{\frac{(x-\nu)_+}{\nu}}-h_{\nu,\zeta}(x)\right|
    &\leq
    \sqrt{\left|
        \frac{(x-\nu)_+}{\nu}-h_{\nu,\zeta}(x)^2
    \right|}\\
    &\leq
    \frac{\sqrt[4]{3}\sqrt{C_\zeta}}{\sqrt[4]{\nu}}
    \longrightarrow0.
\end{aligned}
\]
If $h_{\nu,\zeta}(x)>1$, we have
\[
\begin{aligned}
    \left|\sqrt{\frac{(x-\nu)_+}{\nu}}-h_{\nu,\zeta}(x)\right|
    &=
    \frac{\left|\frac{(x-\nu)_+}{\nu}-h_{\nu,\zeta}(x)^2\right|}
    {\sqrt{\frac{(x-\nu)_+}{\nu}}+h_{\nu,\zeta}(x)}\\
    &\leq
    \frac{C_\zeta}{\sqrt{\nu}}
    \frac{\sqrt{1+2h_{\nu,\zeta}(x)^2}}{h_{\nu,\zeta}(x)}\\
    &\leq
    \frac{\sqrt{3}C_\zeta}{\sqrt{\nu}}
    \longrightarrow0.
\end{aligned}
\]
The bounds in all three cases are independent of $x$. Taking their
maximum therefore gives
\[
\sup_{x\geq0}
\left|
h_{\nu,\zeta}(x)
-
\sqrt{\frac{(x-\nu)_+}{\nu}}
\right| \leq \max
\left\{\frac{\sqrt[4]{3}\sqrt{C_\zeta}}{\sqrt[4]{\nu}},\frac{\sqrt{3}C_\zeta}{\sqrt{\nu}}\right\}
\longrightarrow0.
\]
This proves the uniform convergence.
\end{proof}

\subsection{Meta-analysis with treatment-control studies}
\label{app:proofs-REMA}

\subsubsection{Asymptotic properties of REMA}

\begin{proof}[Proof of Theorem~\ref{thm:dl-REMA-mean-clt}]
We first establish the rate of $\hat\sigma_{\textup{B}}^2$ in part (a), then introduce the oracle estimator and prove the central limit theorem in part (b).

Assumption~\ref{ass:dl-regularity} bounds the sampling variances away from
zero and, through the fourth-moment bound, from above. Consequently,
\[
    \sum_{j=1}^J \sigma_j^{-2}
    -\frac{\sum_{j=1}^J \sigma_j^{-4}}{\sum_{j=1}^J \sigma_j^{-2}}
    \asymp J.
\]
Substituting \eqref{eq:fixed-effect-estimator} into $Q$ gives
\[
    Q
    =
    \sum_{j=1}^J \sigma_j^{-2}(\hat\tau_j-\tau)^2
    -
    \frac{\left\{\sum_{j=1}^J \sigma_j^{-2}(\hat\tau_j-\tau)\right\}^2}{\sum_{j=1}^J \sigma_j^{-2}}.
\]
Therefore
\[
\begin{aligned}
    \E(Q)
    &=
    \sum_{j=1}^J \sigma_j^{-2}(\sigma_j^2+\sigma_{\textup{B}}^2)
    -
    \frac{\sum_{j=1}^J \sigma_j^{-4}(\sigma_j^2+\sigma_{\textup{B}}^2)}{\sum_{j=1}^J \sigma_j^{-2}}  \\
    &=
    (J-1)+\left\{\sum_{j=1}^J \sigma_j^{-2}-\frac{\sum_{j=1}^J \sigma_j^{-4}}{\sum_{j=1}^J \sigma_j^{-2}}\right\}\sigma_{\textup{B}}^2 .
\end{aligned}
\]

Independence and the uniform fourth-moment bound in
Assumption~\ref{ass:dl-regularity} imply
\[
    \Var(Q)=O(J),
\]
and hence
\[
    Q-\E(Q)=O_{\PP}(J^{1/2}).
\]
It follows that
\[
    \frac{Q-(J-1)}{\displaystyle\sum_{j=1}^J \sigma_j^{-2}-\frac{\sum_{j=1}^J \sigma_j^{-4}}{\sum_{j=1}^J \sigma_j^{-2}}}-\sigma_{\textup{B}}^2
    =
    \frac{Q-\E(Q)}{\displaystyle\sum_{j=1}^J \sigma_j^{-2}-\frac{\sum_{j=1}^J \sigma_j^{-4}}{\sum_{j=1}^J \sigma_j^{-2}}}
    =
    O_{\PP}(J^{-1/2}).
\]
Finally, the positive-part map $x\mapsto \max(0,x)$ is Lipschitz and
$\sigma_{\textup{B}}^2\geq0$. Thus the positive-part truncation does not change the rate, which proves (a).

To prove part (b), define the oracle precision-weighted estimator
\[
    \hat\tau_{\RE}^{\mathrm{oracle}}
    =
    \sum_{j=1}^J \lambda_{j,\RE}\hat\tau_j,
    \qquad
    \lambda_{j,\RE}
    =
    \frac{(\sigma_j^2+\sigma_{\textup{B}}^2)^{-1}}
    {\sum_{k=1}^J(\sigma_k^2+\sigma_{\textup{B}}^2)^{-1}},
\]
with variance
\[
    \sigma_{\RE}^2
    =
    \left\{
        \sum_{j=1}^J
        (\sigma_j^2+\sigma_{\textup{B}}^2)^{-1}
    \right\}^{-1}.
\]

By (a), the event $|\hat\sigma_{\textup{B}}^2-\sigma_{\textup{B}}^2|\le1$ has probability
tending to one. On this event,
Assumption~\ref{ass:dl-regularity} implies
\[
    \sum_{k=1}^J(\sigma_k^2+\hat\sigma_{\textup{B}}^2)^{-1}\asymp J.
\]
Applying the reciprocal identity twice gives
\[
\begin{aligned}
\hat\tau_{\RE}-\hat\tau_{\RE}^{\mathrm{oracle}}
&=
\frac{1}{
    \sum_{k=1}^J
    (\sigma_k^2+\hat\sigma_{\textup{B}}^2)^{-1}
}
\sum_{j=1}^J
    \frac{\hat\tau_j-\hat\tau_{\RE}^{\mathrm{oracle}}}{\sigma_j^2+\hat\sigma_{\textup{B}}^2} \\
&=
\frac{1}{
    \sum_{k=1}^J
    (\sigma_k^2+\hat\sigma_{\textup{B}}^2)^{-1}
}
\sum_{j=1}^J
\left\{
    \frac{1}{\sigma_j^2+\hat\sigma_{\textup{B}}^2}
    -
    \frac{1}{\sigma_j^2+\sigma_{\textup{B}}^2}
\right\}
\left(
    \hat\tau_j-\hat\tau_{\RE}^{\mathrm{oracle}}
\right) \\
&=
-\frac{\hat\sigma_{\textup{B}}^2-\sigma_{\textup{B}}^2}{
    \sum_{k=1}^J
    (\sigma_k^2+\hat\sigma_{\textup{B}}^2)^{-1}
}
\sum_{j=1}^J
\frac{
    \hat\tau_j-\hat\tau_{\RE}^{\mathrm{oracle}}
}{
    (\sigma_j^2+\sigma_{\textup{B}}^2)
    (\sigma_j^2+\hat\sigma_{\textup{B}}^2)
} \\
&=
-\frac{\hat\sigma_{\textup{B}}^2-\sigma_{\textup{B}}^2}{
    \sum_{k=1}^J
    (\sigma_k^2+\hat\sigma_{\textup{B}}^2)^{-1}
}
\sum_{j=1}^J
\frac{
    \hat\tau_j-\hat\tau_{\RE}^{\mathrm{oracle}}
}{
    (\sigma_j^2+\sigma_{\textup{B}}^2)^2
} \\
&\quad+
\frac{(\hat\sigma_{\textup{B}}^2-\sigma_{\textup{B}}^2)^2}{
    \sum_{k=1}^J
    (\sigma_k^2+\hat\sigma_{\textup{B}}^2)^{-1}
}
\sum_{j=1}^J
\frac{
    \hat\tau_j-\hat\tau_{\RE}^{\mathrm{oracle}}
}{
    (\sigma_j^2+\sigma_{\textup{B}}^2)^2
    (\sigma_j^2+\hat\sigma_{\textup{B}}^2)
}.
\end{aligned}
\]
To control the first term, Chebyshev's inequality gives
\[
    \sum_{j=1}^J
    \frac{\hat\tau_j-\hat\tau_{\RE}^{\mathrm{oracle}}}
    {(\sigma_j^2+\sigma_{\textup{B}}^2)^2}=
    \sum_{j=1}^J
    \left\{
        \frac{1}{(\sigma_j^2+\sigma_{\textup{B}}^2)^2}
        -
        \frac{
            (\sigma_j^2+\sigma_{\textup{B}}^2)^{-1}
            \sum_{k=1}^J(\sigma_k^2+\sigma_{\textup{B}}^2)^{-2}
        }{
            \sum_{k=1}^J(\sigma_k^2+\sigma_{\textup{B}}^2)^{-1}
        }
    \right\}
    (\hat\tau_j-\tau) =O_{\PP}(J^{1/2}),
\]
since the coefficients in braces are fixed and uniformly bounded. For the second term, Chebyshev's inequality gives
\[
    \frac{1}{J}\sum_{j=1}^J|\hat\tau_j-\hat\tau_{\RE}^{\mathrm{oracle}}|\leq |\hat\tau_{\RE}^{\mathrm{oracle}}-\tau|+ \frac{1}{J}\sum_{j=1}^J|\hat\tau_j-\tau|= O_{\PP}(1),
\]
and thus \[
\frac{1}{
    \sum_{k=1}^J
    (\sigma_k^2+\hat\sigma_{\textup{B}}^2)^{-1}
}
\sum_{j=1}^J
\frac{
    \hat\tau_j-\hat\tau_{\RE}^{\mathrm{oracle}}
}{
    (\sigma_j^2+\sigma_{\textup{B}}^2)^2
    (\sigma_j^2+\hat\sigma_{\textup{B}}^2)
} =O_{\PP}(1).
\]
Consequently,
\begin{equation}
\label{eq:tau-RE-converge-to-oracle}
    \hat\tau_{\RE}-\hat\tau_{\RE}^{\mathrm{oracle}}
    =
    O_{\PP}\!\left\{
        \frac{|\hat\sigma_{\textup{B}}^2-\sigma_{\textup{B}}^2|}{J^{1/2}}
        +(\hat\sigma_{\textup{B}}^2-\sigma_{\textup{B}}^2)^2
    \right\}
    =
    O_{\PP}(J^{-1})
    =
    o_{\PP}(J^{-1/2}).
\end{equation}

By definition,
\[
    \frac{\hat\sigma_{\RE}^2}{\sigma_{\RE}^2}
    =
    \frac{\sum_{j=1}^J(\sigma_j^2+\sigma_{\textup{B}}^2)^{-1}}
    {\sum_{j=1}^J(\sigma_j^2+\hat\sigma_{\textup{B}}^2)^{-1}}.
\]
Hence,
\begin{equation}
\label{eq:sigma-RE-converge-to-oracle}
\begin{aligned}
    \left|
        \frac{\hat\sigma_{\RE}^2}{\sigma_{\RE}^2}-1
    \right|
    &\leq
    |\hat\sigma_{\textup{B}}^2-\sigma_{\textup{B}}^2|
    \frac{
        \sum_{j=1}^J
        (\sigma_j^2+\sigma_{\textup{B}}^2)^{-1}
        (\sigma_j^2+\hat\sigma_{\textup{B}}^2)^{-1}
    }{
        \sum_{j=1}^J(\sigma_j^2+\hat\sigma_{\textup{B}}^2)^{-1}
    } \leq
    O(|\hat\sigma_{\textup{B}}^2-\sigma_{\textup{B}}^2|)
    \overset{\PP}{\to}0.
\end{aligned}
\end{equation}

Under Assumption~\ref{ass:REMA-without-normality}, 
\[
    \sigma_{\RE}^2
    =\sum_{j=1}^J\lambda_{j,\RE}^2(\sigma_j^2+\sigma_{\textup{B}}^2)
    \asymp J^{-1}.
\]
The Lyapunov condition therefore holds:
\[
    \frac{\sum_{j=1}^J
        \E|\lambda_{j,\RE}(\hat\tau_j-\tau)|^4}
        {\sigma_{\RE}^4}
    \leq
    \frac{M\sum_{j=1}^J\lambda_{j,\RE}^4}{\sigma_{\RE}^4}
    =O(J^{-1})\longrightarrow0.
\]
By the Lyapunov central limit theorem,
\[
    \frac{\hat\tau_{\RE}^{\mathrm{oracle}}-\tau}{\sigma_{\RE}}
    \Rightarrow N(0,1).
\]
Therefore, \eqref{eq:tau-RE-converge-to-oracle} and \eqref{eq:sigma-RE-converge-to-oracle} with Slutsky's theorem prove (b).
\end{proof}

\subsubsection{FEMA estimator under REMA}

\begin{proof}[Proof of Theorem~\ref{thm:FEMA-under-REMA}]
The FEMA estimator is a weighted sum of independent study estimates.
Under Assumption~\ref{ass:REMA-without-normality},
\[
    \max_{1\leq j\leq J}\lambda_{j,\FE}=O(J^{-1}),
    \qquad
    \sigma_{\FE\dagger}^2
    =\sum_{j=1}^J\lambda_{j,\FE}^2(\sigma_j^2+\sigma_{\textup{B}}^2)
    \asymp J^{-1}.
\]
The Lyapunov condition therefore holds:
\[
    \frac{\sum_{j=1}^J
        \E|\lambda_{j,\FE}(\hat\tau_j-\tau)|^4}
        { \sigma_{\FE\dagger}^4}
    \leq
    \frac{M\sum_{j=1}^J\lambda_{j,\FE}^4}
        {\sigma_{\FE\dagger}^4}
    =O(J^{-1})\longrightarrow0.
\]
By the Lyapunov central limit theorem,
\[
    \frac{\hat\tau_{\FE}-\tau}{\sigma_{\FE\dagger}}
    \Rightarrow N(0,1).
\]
By Theorem~\ref{thm:dl-REMA-mean-clt}(a),
\[
\begin{aligned}
    \left|
        \frac{\hat{\sigma}_{\FE\dagger}^2 }
             { \sigma_{\FE\dagger}^2 }-1
    \right|
    &=
    \frac{|\hat\sigma_{\textup{B}}^2-\sigma_{\textup{B}}^2|
        \sum_{j=1}^J\lambda_{j,\FE}^2}
        { \sigma_{\FE\dagger}^2 } = O(|\hat\sigma_{\textup{B}}^2-\sigma_{\textup{B}}^2|)
    \overset{\PP}{\to}0,
\end{aligned}
\]
which completes the proof by Slutsky's theorem.
\end{proof}

\subsection{BAMA}
\label{app:proofs-treatment-control}

\subsubsection{Confidence intervals and the maximum bias for significance}

\begin{proof}[Proof of Theorem~\ref{thm:treatment-control-ci}]
Under Assumption~\ref{ass:BAMA-model},
\[
    \hat\tau_{\FE}
    \sim
    N\left\{
    \sum_{j=1}^J\lambda_{j,\FE}\tau_j,
    \sum_{j=1}^J\lambda_{j,\FE}^2\sigma_j^2
    \right\}
    =N(\tau+\Delta_{\FE},\sigma_{\FE}^2),
\]
where
\[
    |\Delta_{\FE}|
    =
    \left|\sum_{j=1}^J\lambda_{j,\FE}\Delta_j\right|
    \leq
    \sum_{j=1}^J\lambda_{j,\FE}|\Delta_j|
    \leq
    b\sqrt J\sigma_{\FE}\sum_{j=1}^J\lambda_{j,\FE}
    =b\sqrt J\sigma_{\FE},
\]
which can be attained by taking $\Delta_j=b\sqrt J\sigma_{\FE}$ for all $j$ or $\Delta_j=-b\sqrt J\sigma_{\FE}$ for all $j$.

For a fixed bias vector $\bm{\Delta}$,
\[
    \frac{\hat\tau_{\FE}-\tau}{\sigma_{\FE}}
    \sim
    N\left(\frac{\Delta_{\FE}}{\sigma_{\FE}},1\right).
\]
Therefore the coverage probability of the centered interval with radius $L$ is
\[
    \Phi\left(L-\frac{\Delta_{\FE}}{\sigma_{\FE}}\right)
    -
    \Phi\left(-L-\frac{\Delta_{\FE}}{\sigma_{\FE}}\right),
\]
which decreases in $|\Delta_{\FE}/\sigma_{\FE}|$, whose supremum is $\sqrt J\,b$ over $\bm{\Delta}\in\cD_b$.
Thus the minimum coverage equals
\[
    \Phi(L-\sqrt Jb)-\Phi(-L-
    \sqrt Jb).
\]
Choosing $L=L_\zeta(\sqrt J\,b)$ makes this minimum coverage exactly
$1-\zeta$. Since the supremum is attainable, any smaller fixed radius fails to
achieve the required coverage, which proves (a).

The confidence interval in (a) excludes zero if and only if
\[
    |t_{\FE}|>L_\zeta(\sqrt Jb).
\]
If $|t_{\FE}|\leq z_{1-\zeta/2}=L_\zeta(0)$, the interval already
contains zero at $b=0$, and $\overline{b}(\zeta)=0$. If
$|t_{\FE}|>z_{1-\zeta/2}$, continuity and strict monotonicity of
$L_\zeta$ give a unique positive boundary value satisfying
\[
    L_\zeta\{\sqrt J \overline{b}(\zeta)\}
    =
    |t_{\FE}|.
\]
By definition of $L_\zeta$, we have
\[
    \Phi\left\{|t_{\FE}|
    -\sqrt J \overline{b}(\zeta)\right\}
    -
    \Phi\left\{-|t_{\FE}|
    -\sqrt J \overline{b}(\zeta)\right\}
    =
    1-\zeta,
\]
which proves (b).

Hence, by Lemma~\ref{lem:hzeta-large-signal}, as $J\to\infty$
\[
    \left|
        \overline{b}(\zeta)
        -
        \frac{|t_{\FE}|}{\sqrt J}
    \right|= \left|
        \frac{1}{\sqrt J}h_\zeta(|t_{\FE}|)
        -
        \frac{|t_{\FE}|}{\sqrt J}
    \right|
    \leq
    \frac{z_{1-\zeta/2}}{\sqrt J}
    \longrightarrow0,
\]
which proves (c).
\end{proof}

\subsubsection{Asymptotic validity without normality}
\begin{proof}[Proof of Proposition~\ref{prop:treatment-control-nongaussian}]
We first obtain a normal approximation that is uniform over the allowed
biases. The fourth-moment bound implies, for every
$\bm{\Delta}\in\cD_b$,
\[
    \E|\hat\tau_j-\tau-\Delta_j|^3\leq M_b^{3/4},
    \qquad
    c_\sigma^2\leq\sigma_j^2
    =\E\{(\hat\tau_j-\tau-\Delta_j)^2\}
    \leq M_b^{1/2}.
\]
Hence, the FEMA weights and variance satisfy
\[
    \max_{1\leq j\leq J}\lambda_{j,\FE}=O(J^{-1}),
    \qquad
    \sigma_{\FE}^2\asymp J^{-1}.
\]
Since
\[
    \hat\tau_{\FE}-\tau-\Delta_{\FE}
    =\sum_{j=1}^J\lambda_{j,\FE}(\hat\tau_j-\tau-\Delta_j),
\]
the Berry--Esseen inequality gives
\[
\begin{aligned}
    &\sup_{\bm{\Delta}\in\cD_b}\sup_{x\in\R}
    \left|
        \PP\left\{
            \frac{\hat\tau_{\FE}-\tau-\Delta_{\FE}}
                 {\sigma_{\FE}}\leq x
        \right\}-\Phi(x)
    \right|\lesssim
    \frac{C}{\sigma_{\FE}^3}
    \sup_{\bm{\Delta}\in\cD_b}
    \sum_{j=1}^J\lambda_{j,\FE}^3
        \E|\hat\tau_j-\tau-\Delta_j|^3=O(J^{-1/2}).
\end{aligned}
\]

We now use this approximation to calculate coverage. Since $\Phi$ is
continuous, the same bound controls probabilities of closed intervals,
with at most twice the approximation error. Hence,
\[
\begin{aligned}
    \PP\left\{
        |\hat\tau_{\FE}-\tau|
        \leq\sigma_{\FE}L_\zeta(\sqrt J\,b)
    \right\}
    &={}
    \Phi\left\{
        L_\zeta(\sqrt J\,b)-\frac{\Delta_{\FE}}{\sigma_{\FE}}
    \right\} \\
    &\quad-
    \Phi\left\{
        -L_\zeta(\sqrt J\,b)-\frac{\Delta_{\FE}}{\sigma_{\FE}}
    \right\}+O(J^{-1/2}),
\end{aligned}
\]
uniformly over $\bm{\Delta}\in\cD_b$. This remains true even though the
interval endpoints vary with $J$, because the normal approximation is
uniform in $x$.

By the definition of $\cD_b$ and the nonnegative FEMA weights,
\[
    \left|\frac{\Delta_{\FE}}{\sigma_{\FE}}\right|
    =\left|
        \frac{\sum_{j=1}^J\lambda_{j,\FE}\Delta_j}{\sigma_{\FE}}
    \right|
    \leq\sqrt J\,b.
\]
We attain equality by taking $\Delta_j=\sqrt J\,b\,\sigma_{\FE}$
for every $j$. Since
$\Phi(L-x)-\Phi(-L-x)$ decreases with $|x|$
for $L>0$, taking the infimum in the coverage approximation gives
\[
\begin{aligned}
    \inf_{\bm{\Delta}\in\cD_b}
    \PP\left\{
        |\hat\tau_{\FE}-\tau|
        \leq\sigma_{\FE}L_\zeta(\sqrt J\,b)
    \right\}
    &={}
    \Phi\left\{L_\zeta(\sqrt J\,b)-\sqrt J\,b\right\} \\
    &\quad-
    \Phi\left\{-L_\zeta(\sqrt J\,b)-\sqrt J\,b\right\}
    +O(J^{-1/2}) \\
    &=1-\zeta+O(J^{-1/2}),
\end{aligned}
\]
by the definition of $L_\zeta$, which completes the proof.
\end{proof}

\subsubsection{Model checking of BAMA and calibrating the maximum bias for significance}

\begin{proof}[Proof of Theorem~\ref{thm:treatment-control-Q}]
For each fixed bias vector, $Q$ has a noncentral chi-square distribution.
We maximize its noncentrality over $\cD_b$ to obtain a valid
upper-tail test. Specifically, 
\[
    Q\sim\chi^2_{J-1}\left\{
        \sum_{j=1}^J\sigma_j^{-2}
        (\Delta_j-\Delta_{\FE})^2
    \right\}.
\]
The noncentrality parameter is a convex function of $\bm{\Delta}$, so it attains its maximum over $\cD_b$ at a vertex
$\Delta_j=b\sqrt J\,\sigma_{\FE}s_j$, where
$s_j\in\{-1,1\}$ for all $j=1,\ldots,J$. Hence
\[
\begin{aligned}
    &\quad\sup_{\bm{\Delta}\in\cD_b}
    \sum_{j=1}^J\sigma_j^{-2}(\Delta_j-\Delta_{\FE})^2\\
    &=Jb^2\sigma_{\FE}^2
    \max_{{s_1,\ldots,s_J\in\{-1,1\}}}
    \left\{
        \sum_{j=1}^J\sigma_j^{-2}
        -\frac{(\sum_{j=1}^J\sigma_j^{-2}s_j)^2}
        {\sum_{j=1}^J\sigma_j^{-2}}
    \right\}\\
    &=Jb^2\left\{1-
    \min_{{s_1,\ldots,s_J\in\{-1,1\}}}
    \left(\sum_{j=1}^J\lambda_{j,\FE}s_j\right)^2\right\}
    =\Lambda_b.
\end{aligned}
\]
The upper-tail probability of a noncentral $\chi^2$ variable is
nondecreasing in its noncentrality parameter. Therefore, for every
$\alpha\in(0,1)$ and every $\bm{\Delta}\in\cD_b$,
\[
    \PP\left\{
        1-\operatorname{pchisq}
        (Q;\mathrm{df}=J-1,\mathrm{ncp}=\Lambda_b)
        \leq\alpha
    \right\}
    \leq\alpha,
\]
which proves (a).

Let
$S_1,\ldots,S_J$ be independent Rademacher random variables. We have
\[
\begin{aligned}
    \min_{s\in\{-1,1\}^J}
    \left(\sum_{j=1}^J\lambda_{j,\FE}s_j\right)^2
    &\leq
    \E\left(\sum_{j=1}^J\lambda_{j,\FE}S_j\right)^2 =\sum_{j=1}^J\lambda_{j,\FE}^2
    \leq\max_{1\leq j\leq J}\lambda_{j,\FE}
    \longrightarrow0.
\end{aligned}
\]
Consequently, $\Lambda_1/J\to1$.

By definition and the assumed boundedness,
\begin{equation}
\label{eq:excess-Q-over-J-basic}
    \sqrt{
        \frac{\{Q-(J-1)\}_+}{J}
    }
    =
    \frac{\widehat\sigma_{\textup{B}}}{\sqrt J\,\sigma_{\FE}}
    \sqrt{1-\sum_{j=1}^J\lambda_{j,\FE}^2}
    =\frac{\widehat\sigma_{\textup{B}}}{\sqrt J\,\sigma_{\FE}}+o_{\PP}(1).
\end{equation}
By definition and $\Lambda_1/J\to1$, Lemma~\ref{lem:nchisq-uniform-inversion} gives
\[
    \underline{b}(\zeta)
    =
    \sqrt{\frac{J-1}{\Lambda_1}}
    h_{J-1,\zeta}(Q) = \sqrt{
        \frac{\{Q-(J-1)\}_+}{\Lambda_1}
    } + o_{\PP}(1),
\]
and
\[
\begin{aligned}
    &\sqrt{
        \frac{\{Q-(J-1)\}_+}{\Lambda_1}
    }
    -
    \sqrt{
        \frac{\{Q-(J-1)\}_+}{J}
    }=
    \sqrt{
        \frac{\{Q-(J-1)\}_+}{J}
    }
    \left\{\sqrt{\frac{J}{\Lambda_1}}-1\right\}
    =o_{\PP}(1).
\end{aligned}
\]
This completes the proof by \eqref{eq:excess-Q-over-J-basic}.
\end{proof}

\begin{proof}[Proof of Theorem~\ref{thm::empty-interval}]
We will show that under
Assumption~\ref{ass:BAMA-model} with 
\[
    \max_{1\leq j\leq J}\lambda_{j,\FE}\longrightarrow 0, \quad
    J\to\infty ,
\]
we have
\[
    \PP(\mathcal E_1\mid\mathcal E_2)\longrightarrow 1,
\]
where the events are defined as 
\[
\begin{aligned}
\mathcal E_1
&=\{[\underline b(\zeta),\overline b(\zeta)) = \varnothing\},\\
\mathcal E_2
&=\left\{
    p_Q(0)<\zeta,\;
    \left|
        \frac{\hat\tau_{\FE}}
        {\hat{\sigma}_{\FE\dagger}}
    \right|
    \leq z_{1-\zeta/2}
\right\}.
\end{aligned}
\]

By definition,
\[
    \hat\sigma_{\textup{B}}^2
    =
    \frac{\sigma_{\FE}^2}
    {1-\sum_{j=1}^J\lambda_{j,\FE}^2}
    \{Q-(J-1)\}_+.
\]
Therefore, 
\[
    \frac{\hat{\sigma}_{\FE\dagger}^2}
    {\sigma_{\FE}^2}
    =
    1+
    \frac{\sum_{j=1}^J\lambda_{j,\FE}^2}
    {1-\sum_{j=1}^J\lambda_{j,\FE}^2}
    \{Q-(J-1)\}_+.
\]
Moreover,
\[
    \sum_{j=1}^J\lambda_{j,\FE}^2
    \leq
    \max_{1\leq j\leq J}\lambda_{j,\FE}
    \longrightarrow0.
\]

On the conditioning event, $p_Q(0)<\zeta$, so
$\underline b(\zeta)>0$ and $Q>J-1$. If
$|t_{\FE}|\leq z_{1-\zeta/2}$, then
$\overline b(\zeta)=0$ by
Theorem~\ref{thm:treatment-control-bvalue}, and the desired ordering
already holds. It remains to consider
$|t_{\FE}|>z_{1-\zeta/2}$, hence
\begin{equation}
\label{eq:threshold-ordering-t-bound}
\begin{aligned}
    0<|t_{\FE}|^2-z_{1-\zeta/2}^2\leq
    z_{1-\zeta/2}^2
    \frac{\sum_{j=1}^J\lambda_{j,\FE}^2}
    {1-\sum_{j=1}^J\lambda_{j,\FE}^2}
    \{Q-(J-1)\}.
\end{aligned}
\end{equation}

We next convert this inequality into a bound on the maximum bias for significance. A Taylor
expansion of the defining equation for $L_\zeta$ at zero gives
\[
    L_\zeta(x)^2-z_{1-\zeta/2}^2
    =z_{1-\zeta/2}^2x^2+O(x^4)
    \qquad (x\to0).
\]
Thus,
$h_\zeta(s)^2=O\{s^2-z_{1-\zeta/2}^2\}$ as
$s\downarrow z_{1-\zeta/2}$. Lemma~\ref{lem:hzeta-large-signal}
gives $0\leq s-h_\zeta(s)\leq z_{1-\zeta/2}$, which gives the same
bound for large $s$; continuity gives it on the remaining compact
interval. Therefore,
\[
    h_\zeta(s)^2
    =O\{s^2-z_{1-\zeta/2}^2\},
    \qquad s>z_{1-\zeta/2}.
\]
By definition and
\eqref{eq:threshold-ordering-t-bound},
\begin{equation}
\label{eq:threshold-ordering-bvalue-bound}
\begin{aligned}
    J\overline b(\zeta)^2
    =h_\zeta(|t_{\FE}|)^2=O\left[
        \left(\sum_{j=1}^J\lambda_{j,\FE}^2\right)
        \{Q-(J-1)\}
    \right].
\end{aligned}
\end{equation}

Suppose now that
$\overline b(\zeta)\geq\underline b(\zeta)$. Since $p_Q(b)$ is
nondecreasing in $b$,
\[
    p_Q\{\overline b(\zeta)\}\geq\zeta.
\]
Thus $Q$ is larger than the $(1-\zeta)$ quantile of the central
$\chi^2_{J-1}$ distribution, but no larger than the corresponding
quantile of the noncentral $\chi^2_{J-1}$ distribution with
noncentrality parameter $\Lambda_{\overline b(\zeta)}$. By
Theorem~\ref{thm:treatment-control-Q}(a) and
\eqref{eq:threshold-ordering-bvalue-bound},
\begin{equation}
\label{eq:threshold-ordering-lambda-bound}
\begin{aligned}
    \Lambda_{\overline b(\zeta)}
    \leq J\overline b(\zeta)^2
    =O\left[
        \left(\sum_{j=1}^J\lambda_{j,\FE}^2\right)
        \{Q-(J-1)\}
    \right].
\end{aligned}
\end{equation}

A noncentral $\chi^2_{J-1}$ variable with noncentrality parameter
$\Lambda_{\overline b(\zeta)}$ has mean
$J-1+\Lambda_{\overline b(\zeta)}$ and variance
$2\{J-1+2\Lambda_{\overline b(\zeta)}\}$. Chebyshev's inequality and
the preceding quantile comparison therefore give
\[
    Q-(J-1)
    =O\left\{
        \Lambda_{\overline b(\zeta)}
        +\sqrt{J+\Lambda_{\overline b(\zeta)}}
    \right\}.
\]
Using \eqref{eq:threshold-ordering-lambda-bound}, moving the term
proportional to $Q-(J-1)$ to the left, and then squaring give
\[
    \{Q-(J-1)\}^2
    =O\left[
        J+
        \left(\sum_{j=1}^J\lambda_{j,\FE}^2\right)
        \{Q-(J-1)\}
    \right].
\]
Solving this quadratic inequality yields
\[
    Q-(J-1)=O(\sqrt J).
\]
Substitution into \eqref{eq:threshold-ordering-lambda-bound} then gives
\begin{equation}
\label{eq:threshold-ordering-lambda-small}
    \Lambda_{\overline b(\zeta)}
    =O\left(
        \sqrt J\sum_{j=1}^J\lambda_{j,\FE}^2
    \right)
    =o(\sqrt J).
\end{equation}

Recall the quantile notation $q_{1-\zeta}(\nu,\lambda)$ used in the
proof of Lemma~\ref{lem:nchisq-uniform-inversion}. For
$0\leq\lambda\leq\Lambda_{\overline b(\zeta)}$,
Lemma~\ref{lem:nchisq-uniform-inversion} and
\eqref{eq:threshold-ordering-lambda-small} imply
\[
    q_{1-\zeta}(J-1,\lambda)=O(J)
\]
uniformly, since
\[
    h_{J-1,\zeta}\{q_{1-\zeta}(J-1,\lambda)\}
    =\sqrt{\frac{\lambda}{J-1}},
\]
and Lemma~\ref{lem:nchisq-uniform-inversion} gives
\[
    \sqrt{
        \frac{
            \{q_{1-\zeta}(J-1,\lambda)-(J-1)\}_+
        }{J-1}
    }
    =
    \sqrt{\frac{\lambda}{J-1}}+o(1)
    =o(1)
\]
uniformly over this range.

If $\overline b(\zeta)\geq\underline b(\zeta)$ on the conditioning
event, then
\[
    p_Q(0)<\zeta
    \leq p_Q\{\overline b(\zeta)\},
\]
and therefore
\[
    q_{1-\zeta}(J-1,0)
    <Q
    \leq
    q_{1-\zeta}\{J-1,\Lambda_{\overline b(\zeta)}\}.
\]

We then show that, for every deterministic sequence
$\lambda_J=o(\sqrt J)$,
\[
\PP\left\{
    q_{1-\zeta}(J-1,0)<Q
    \leq q_{1-\zeta}(J-1,\lambda_J)
\right\}
=o(1)
\]
uniformly over $\bm{\Delta}\in\cD_b$. Let
$Z_1,\ldots,Z_{J-1}$ be independent standard normal variables and,
on the same probability space, define
\[
\begin{aligned}
    Q_0
    &=
    Z_1^2+\sum_{j=2}^{J-1}Z_j^2,\\
    Q_1
    &=
    (Z_1+\sqrt{\lambda_J})^2
    +\sum_{j=2}^{J-1}Z_j^2,\\
    Q
    &=
    \left[
        Z_1+
        \left\{
            \sum_{j=1}^J
            \sigma_j^{-2}
            (\Delta_j-\Delta_{\FE})^2
        \right\}^{1/2}
    \right]^2
    +\sum_{j=2}^{J-1}Z_j^2.
\end{aligned}
\]
Then
\[
    Q_0\sim\chi^2_{J-1},
    \qquad
    Q_1\sim\chi^2_{J-1}(\lambda_J),
\]
and $Q$ has the distribution in
Assumption~\ref{ass:BAMA-model}.

Since
\[
    Q_1-Q_0
    =
    2\sqrt{\lambda_J}Z_1+\lambda_J = o_{\PP}(\sqrt{J}),
\]
and
\[
    \frac{Q_0-(J-1)}{\sqrt{2(J-1)}}
    \overset{d}{\longrightarrow}N(0,1),
\]
Slutsky's theorem gives
\[
    \frac{Q_1-(J-1)}{\sqrt{2(J-1)}}
    \overset{d}{\longrightarrow}N(0,1).
\]
Since the standard normal distribution function is continuous and
strictly increasing,
\[
\begin{aligned}
    \frac{
        q_{1-\zeta}(J-1,0)-(J-1)
    }{
        \sqrt{2(J-1)}
    }
    &\longrightarrow z_{1-\zeta},\\
    \frac{
        q_{1-\zeta}(J-1,\lambda_J)-(J-1)
    }{
        \sqrt{2(J-1)}
    }
    &\longrightarrow z_{1-\zeta}.
\end{aligned}
\]
Consequently,
\[
    q_{1-\zeta}(J-1,\lambda_J)
    -
    q_{1-\zeta}(J-1,0)
    =
    o(\sqrt J).
\]

Next, conditional on $Z_1$,
\[
\begin{aligned}
    &\PP\left\{
        q_{1-\zeta}(J-1,0)<Q
        \leq q_{1-\zeta}(J-1,\lambda_J)
        \;\middle|\; Z_1
    \right\}\\
    &\quad\leq
    \sup_{u\in\R}
    \PP\left\{
        u<\sum_{j=2}^{J-1}Z_j^2
        \leq
        u+
        q_{1-\zeta}(J-1,\lambda_J)
        -
        q_{1-\zeta}(J-1,0)
    \right\}.
\end{aligned}
\]
The bound on the right does not depend on $\bm{\Delta}$. Moreover,
\[
    \frac{
        \sum_{j=2}^{J-1}Z_j^2-(J-2)
    }{
        \sqrt{2(J-2)}
    }
    \overset{d}{\longrightarrow}N(0,1).
\]
Therefore,
\[
\begin{aligned}
    &\sup_{u\in\R}
    \PP\left\{
        u<\sum_{j=2}^{J-1}Z_j^2
        \leq
        u+
        q_{1-\zeta}(J-1,\lambda_J)
        -
        q_{1-\zeta}(J-1,0)
    \right\}\\
    \leq&
    2\sup_{t\in\R}
    \left|
        \PP\left\{
            \frac{
                \sum_{j=2}^{J-1}Z_j^2-(J-2)
            }{
                \sqrt{2(J-2)}
            }
            \leq t
        \right\}
        -
        \Phi(t)
    \right|\\
    &+
    \sup_{t\in\R}
    \left[
        \Phi\left\{
            t+
            \frac{
                q_{1-\zeta}(J-1,\lambda_J)
                -
                q_{1-\zeta}(J-1,0)
            }{
                \sqrt{2(J-2)}
            }
        \right\}
        -
        \Phi(t)
    \right]\\
    =&o(1),
\end{aligned}
\]
since $\Phi$ is uniformly continuous. Taking expectations over $Z_1$
therefore gives
\[
    \sup_{\bm{\Delta}\in\cD_b}
    \PP\left\{
        q_{1-\zeta}(J-1,0)<Q
        \leq q_{1-\zeta}(J-1,\lambda_J)
    \right\}
    =o(1).
\]

Finally, by \eqref{eq:threshold-ordering-lambda-small}, we can choose a
deterministic sequence $\lambda_J=o(\sqrt J)$, independent of
$\bm{\Delta}$, such that
$\Lambda_{\overline b(\zeta)}\leq\lambda_J$ on the conditioning
event whenever $\overline b(\zeta)\geq\underline b(\zeta)$.
On this event, monotonicity of the noncentral chi-square quantile gives
\[
    q_{1-\zeta}(J-1,0)<Q
    \leq q_{1-\zeta}\{J-1,\Lambda_{\overline b(\zeta)}\}
    \leq q_{1-\zeta}(J-1,\lambda_J).
\]

On $p_Q(0)<\zeta$, nonrejection by the REMA-adjusted Wald test is the
event
\[
    |t_{\FE}|
    \leq
    z_{1-\zeta/2}
    \left[
        1+
        \frac{\sum_{j=1}^J\lambda_{j,\FE}^2}
        {1-\sum_{j=1}^J\lambda_{j,\FE}^2}
        \{Q-(J-1)\}_+
    \right]^{1/2},
\]
where the right-hand side is nondecreasing in $Q$. Conditional on $Q$, the probability of nonrejection by the REMA-adjusted Wald test is nondecreasing in $Q$, because
$t_{\FE}$ is independent of $Q$ under Assumption~\ref{ass:BAMA-model}, and the acceptance interval
expands with $Q$. Conditioning on this nonrejection therefore
shifts the distribution of $Q$ upward. Thus, for every $x$,
\[
\begin{aligned}
    &\PP\left\{
        Q\leq x
        \;\middle|\;
        p_Q(0)<\zeta,
        \left|
            \frac{\hat\tau_{\FE}}
            {\hat{\sigma}_{\FE\dagger}}
        \right|
        \leq z_{1-\zeta/2}
    \right\}\leq
    \PP\{Q\leq x\mid p_Q(0)<\zeta\}.
\end{aligned}
\]
Stochastic
monotonicity in the noncentrality parameter gives
\[
    \PP\{p_Q(0)<\zeta\}\geq\zeta.
\]
Consequently, since the preceding bounds hold uniformly over
$\bm{\Delta}\in\cD_b$,
\[
\begin{aligned}
    &\,
    \PP\left\{
        \overline b(\zeta)\geq\underline b(\zeta)
        \;\middle|\;
        p_Q(0)<\zeta,
        \left|
            \frac{\hat\tau_{\FE}}
            {\hat{\sigma}_{\FE\dagger}}
        \right|
        \leq z_{1-\zeta/2}
    \right\}\\
    \leq&\,
    \PP\left\{
        Q
        \leq
        q_{1-\zeta}(J-1,\lambda_J)
        \;\middle|\;
        p_Q(0)<\zeta,
        \left|
            \frac{\hat\tau_{\FE}}
            {\hat{\sigma}_{\FE\dagger}}
        \right|
        \leq z_{1-\zeta/2}
    \right\}\\
    \leq&\,
    \PP\left\{
        Q
        \leq
        q_{1-\zeta}(J-1,0)
        +
        o(\sqrt J)
        \;\middle|\;
        p_Q(0)<\zeta,
        \left|
            \frac{\hat\tau_{\FE}}
            {\hat{\sigma}_{\FE\dagger}}
        \right|
        \leq z_{1-\zeta/2}
    \right\}\\
    \leq&\,
    \PP\left\{
        Q
        \leq
        q_{1-\zeta}(J-1,0)
        +
        o(\sqrt J)
        \;\middle|\;
        p_Q(0)<\zeta
    \right\}\\
    =&\,
    \frac{
        \PP\left\{
            q_{1-\zeta}(J-1,0)<Q
            \leq
            q_{1-\zeta}(J-1,0)
            +
            o(\sqrt J)
        \right\}
    }{
        \PP\{p_Q(0)<\zeta\}
    }\\
    \leq&\,
    \frac{1}{\zeta}\,
    \PP\left\{
        q_{1-\zeta}(J-1,0)<Q
        \leq
        q_{1-\zeta}(J-1,0)
        +
            o(\sqrt J)
    \right\}\longrightarrow0,
\end{aligned}
\]
uniformly over $\bm{\Delta}\in\cD_b$, which proves the result.
\end{proof}

\end{document}